\documentclass[%
 reprint,
 amsmath,amssymb,
 aps,
]{revtex4-1}

\usepackage{graphicx}
\usepackage{dcolumn}
\usepackage{bm}
\newcommand{\bra}[1]{\ensuremath{\left\langle#1\right|}}
\newcommand{\ket}[1]{\ensuremath{\left|#1\right\rangle}}
\newcommand{\bracket}[2]{\ensuremath{\left\langle#1 \vphantom{#2}\right| \left. #2 \vphantom{#1}\right\rangle}}
\newcommand{\matrixel}[3]{\ensuremath{\left\langle #1 \vphantom{#2#3} \right| #2 \left| #3 \vphantom{#1#2} \right\rangle}}

\begin{document}
\title{Crystal field, ligand field, and interorbital effects in two-dimensional transition metal dichalcogenides across the periodic table}

\author{Diego Pasquier}
\email{diego.pasquier@epfl.ch}
\author{Oleg V. Yazyev}%
\email{oleg.yazyev@epfl.ch}
\affiliation{Institute of Physics, Ecole Polytechnique F\'{e}d\'{e}rale de Lausanne (EPFL), CH-1015 Lausanne, Switzerland}
\begin{abstract}
Two-dimensional transition metal dichalcogenides (TMDs) exist in two polymorphs, referred to as $1T$ and $1H$, depending on the coordination sphere of the transition metal atom. 
The broken octahedral and trigonal prismatic symmetries lead to different crystal and ligand field splittings of the $d$ electron states, resulting in distinct electronic properties.
In this work, we quantify the crystal and ligand field parameters of two-dimensional TMDs using a Wannier-function approach. 
We adopt the methodology proposed by Scaramucci \textit{et al.} [A. Scaramucci \textit{et al.}, J. Phys.: Condens. Matter 27, 175503 (2015)]. that allows to separate various contributions to the ligand field by choosing different manifolds in the construction of the Wannier functions. 
We discuss the relevance of the crystal and ligand fields in determining the relative stability of the two polymorphs as a function of the filling of the $d$-shell. 
Based on the calculated parameters, we conclude that the ligand field, while leading to a small stabilizing factor  for the $1H$ polymorph in the $d^1$ and $d^2$ TMDs, plays mostly an indirect role and that hybridization between different $d$ orbitals is the dominant feature.
We investigate trends across the periodic table and interpret the variations of the calculated crystal and ligand fields in terms of the change of charge-transfer energy, which allows developing simple chemical intuition.
\end{abstract}
\maketitle
\section{Introduction}\label{sec:intro}
With the advent of two-dimensional (2D) materials \cite{novoselov2005two}, layered transition metal dichalcogenides (TMDs) \cite{wilson1969transition} have gained a great revival of interest due to their rich variety of properties of relevance to both applications and fundamental research \cite{chhowalla_chemistry_2013, wang2012electronics, manzeli_2d_2017, qian_quantum_2014, radisavljevic2011single}.
Two-dimensional TMDs of chemical composition MX$_2$ consist of a triangular lattice of a transition metal atom (M = Mo, W, Re, etc.) sandwiched between two layers of a chalcogen element (X = S, Se, Te).
The local coordination sphere of the transition metal atom can have either trigonal antiprismatic (or distorted octahedral) or trigonal prismatic symmetry, leading to two families of polymorphs, referred to as $1T$ and $1H$, respectively.
The two polymorphs have distinct electronic properties. 
For instance, $1H$-MoS$_2$ is a semiconductor with promising applications \cite{radisavljevic2011single}, $1T$-MoS2 is metallic, and the distorted $1T'$-MoS$_2$ is predicted to be a topological insulator \cite{qian2014quantum}.
In a simplified ionic picture, one assigns formal charges $4+$ and $2-$ to the transition metal and chalcogen ions, respectively \cite{kertesz1984octahedral}.
In such a picture, the formal electronic configuration of the chalcogen atoms is $n_{\mathrm{X}}s^2n_{\mathrm{X}}p^6$ (with $n_{\mathrm{X}}$ = 3, 4, 5 for X = S, Se, Te), while that of the transition metal $M$ is $n_{\mathrm{M}}d^n$ ($n_{\mathrm{M}}=3, 4, 5$), with $n$ depending on the column where $M$ stands in the periodic table ($n=0$ for group IV TMDs such as HfS$_2$, $n=1$ for group V TMDs such as TaS$_2$, and so on).

The electronic properties of the TMDs are therefore governed by the $d$-like bands and their filling \cite{chhowalla_chemistry_2013}. 
The presence of the ligands splits the $d$-electrons levels due to a combination of different effects. 
The crystal field splitting originates from the electrostatic interaction between the orbitals and the ligands, leading to a higher energy for orbitals pointing towards negatively charged ions. 
The ligand field splitting results from hybridization with ligands' orbitals and is expected to be dominant in covalent systems.

According to crystal field theory, in an octahedral environment ($1T$ polymorph), the $d$-shell splits into a low-energy triplet ($t_{2g}$) and a high-energy doublet ($e_g$).
In a trigonal prismatic geometry ($1H$), the low-energy triplet further splits into a doublet and a singlet, usually assumed to be lower in energy.

In the literature, ligand field arguments are often given as simple intuitive starting points to understand several properties of the TMDs.
In particular, a longstanding problem in the field of dichalcogenides is that of the relative stability between the two polymorphs \cite{huisman1971trigonal, kertesz1984octahedral}, which is controlled by the column of the transition metal $M$ in the periodic table, i.e. by the electron filling of the $d$-like bands. 
Indeed, $d^0$ TMDs are found in nature in the $1T$ polymorph, that is also predicted to be energetically more favourable by first-principles calculations.
TMDs with formal occcupation $d^1$-$d^2$ are more stable in the $1H$ polymorph, while the $1T$ polymorph is metastable in a distorted form \cite{duerloo2014structural}.
TMDs with $n=3$ are most stable in a strongly distorted $1T$ phase with $2 \times 2$ periodicity \cite{wilson1969transition, kertesz1984octahedral, whangbo1992analogies, tongay2014monolayer, choi2018origin}, but the $1H$ polymorph is predicted to be lower in energy than the undistorted $1T$ one.
Finally, TMDs in the $d^4-d^6$ range are lower in energy in the $1T$ phase compared to the $1H$. 
Note that for some materials, unrelated pyrite structures are in certain cases the most stable phases \cite{wang2015not}.

A natural explanation for this trend is as follows \cite{yang2017structural}.
For $n<2$  $d$ electrons, the $1H$ phase becomes more and more stable with respect the the $1T$ phase as the filling of the low-energy singlet increases.
On the other hand, for $n>2$ the $1H$ polymorph becomes less and less favourable with increasing the number of electrons because the higher-energy doublet gets filled. 
While being elegant and often invoked in the recent literature \cite{yang2017structural, santosh2015phase}, several authors have argued that it is likely too simplistic because of the complexity of the actual band structure \cite{mattheiss1973, kertesz1984octahedral, isaacs2016}.

The purpose of this paper is twofold.
Firstly, we provide a systematic estimate of crystal and ligand field parameters across the family of materials from first-principles calculations, focusing on the case of monolayers. 
By constructing \textit{ab initio} Wannier tight-binding Hamiltonians for different sets of bands, we estimate the bare crystal field coming from the electrostatic repulsion with the positively charged ions, as well as contributions stemming from hybridization with various ligands' states.
Secondly, in light of the calculated parameters, we discuss the problem of the relative stability of the $1H$ and $1T$ materials as a function of the column of the transition metal $M$ in the periodic table. 
We show that the singlet low-energy state in the $1H$ polymorph is close in energy to the $t_{2g}$ triplet in the $1T$ polymorph, meaning that the ligand field alone plays a minor if any role in determining the relative stability.
However, we also argue that, taking into account interorbital hybridization in the $1H$ case, resulting from nearest-neighbor hoppings between orbitals of different character, the calculated energy diagrams can provide a simple picture for the calculated relative stability of the two phases.

This paper is organized as follows. 
In Sec.~\ref{sec:methodology}, we review the methodology adopted and provide computational details of our calculations.
In Sec.~\ref{sec:tas2}, a detailed study of TaS$_2$ is given as an example.
In Sec.~\ref{sec:stability}, we discuss the relevance of the crystal and ligand field in determining the relative stability of the $1H$ and $1T$ phases, taking again TaS$_2$ as a representative example.
In Sec.~\ref{sec:trend}, we present trends in the calculated parameters across the periodic table.
In Sec.~\ref{sec:previous}, we put our study in perspective with previous work, and Sec. \ref{sec:conclusion} offers conclusions and outlook.
\section{Methodology}\label{sec:methodology}
\subsection{Wannierization and crystal field parameters}
We begin by briefly reviewing the methodolgy proposed in Ref.~\cite{scaramucci2015separating} that we have embraced in order to calculate the crystal field and ligand field parameters. 
Given a set of $n$ isolated bands, one can define a corresponding set of $n$ Wannier functions (WFs) \cite{marzari_maximally_2012} as follows
\begin{equation}
\ket{\mathcal{W}_{R\alpha}} = \frac{1}{\sqrt{N}}\sum_{k, \alpha'} e^{-ikR} U_{\alpha'\alpha}(k)\ket{\psi_{k\alpha'}}\, ,
\end{equation}
where $\ket{\mathcal{W}_{R,\alpha}}$ denotes the Wannier function of character $\alpha$ centered in the lattice site $R$, $N$ is the number of points in the discretized Brillouin zone, $k$ is a pseudomentum, $U_{\alpha'\alpha}(k)$ is the gauge-fixing matrix, and the $\ket{\psi_{k\alpha}}$ are Bloch eigenstates. 
In this work, the Bloch eigenstates $\ket{\psi_{k\alpha}}$ are calculated from density functional theory (DFT) at the level of the generalized gradient approximation (GGA) and correspond to the Kohn-Sham states.
The corresponding Bloch Hamiltonian can be expressed in the basis of the Wannier functions:
\begin{equation}
\begin{split}
H = \sum_{k,\alpha}\epsilon_{k\alpha} \ket{\psi_{k\alpha} }\bra{{\psi_{k\alpha}}} \\ =\sum_{R,R', \alpha, \alpha'} H_{\alpha\alpha'}^{R-R'}\ket{\mathcal{W}_{R\alpha}}\bra{\mathcal{W}_{R'\alpha'}}  \, ,
\end{split}
\end{equation}
where the $\epsilon_{k\alpha}$ are the single-electron eigenenergies (i.e. the Kohn-Sham energies in a standard DFT calculation) and the matrix elements in the Wannier basis $H_{\alpha\alpha}^{RR'}$ can be interpreted as the on-site energies (for $R=R'$ and $\alpha=\alpha'$) and hopping parameters of a tight-binding model.

The construction of Wannier functions contains a high degree of arbitrariness in the choice of the set of bands and the gauge-fixing matrix $U_{\alpha\alpha'}$. 
A common choice for the gauge is that minimizing the spread functional $\Omega$, leading to Maximally Localized Wannier Functions (MLWF) \cite{marzari_maximally_1997, marzari_maximally_2012} :
\begin{equation}
\begin{split}
\frac{\delta \Omega [U]}{\delta U} =0  \, ,\\ 
\Omega=\sum_{\alpha}\left( \matrixel{\mathcal{W}_{0\alpha}}{r^2}{\mathcal{W}_{0\alpha}} - |\matrixel{\mathcal{W}_{0\alpha}}{r}{\mathcal{W}_{0\alpha}}|^2 \right)\, ,
\end{split}
\end{equation}
where $r$ is the position operator.
The advantages of MLWFs are numerous : the constructed Wannier functions are real and atomic-like, the minimization of the spread leads to a minimal overlap between different Wannier functions and therefore optimal interpolation power, and in principle no \textit{a priori} knowledge of the orbital character of the bands is required.
Although MLFWs were adopted as the most useful choice in Ref.~\cite{scaramucci2015separating}, we have found it more convenient for the materials considered to use Projector Wannier Functions (PWFs), defined by using orthogonalized L\"{o}wdin projections of the Bloch eigenstates on hydrogen-like atomic wave functions.
This corresponds to fixing 
\begin{equation}\label{eq:projections}
U_{\alpha'\alpha}(k)= \sum_{\alpha''} (S^{-1/2}(k))_{\alpha''\alpha}\bracket{\psi_{k\alpha'}}{g_{\alpha''}} \, ,
\end{equation} 
where $S$ is the overlap matrix, defined as $S_{\alpha\alpha'}(k)=\sum_n\bracket{g_{\alpha}}{\psi_{kn}}\bracket{\psi_{kn}}{g_{\alpha'}}$, and the $\ket{g_{\alpha}}$ are a set of localized trial orbitals.
This choice allows for a better control of the orbital character of the Wannier functions, that is sometimes lost during the localization procedure.
We shall explain this choice in more details in the next section, and argue that the calculated parameters are consistent with those obtained using MLWFs.
\begin{figure*}[t]
\centering
\includegraphics[width=17cm]{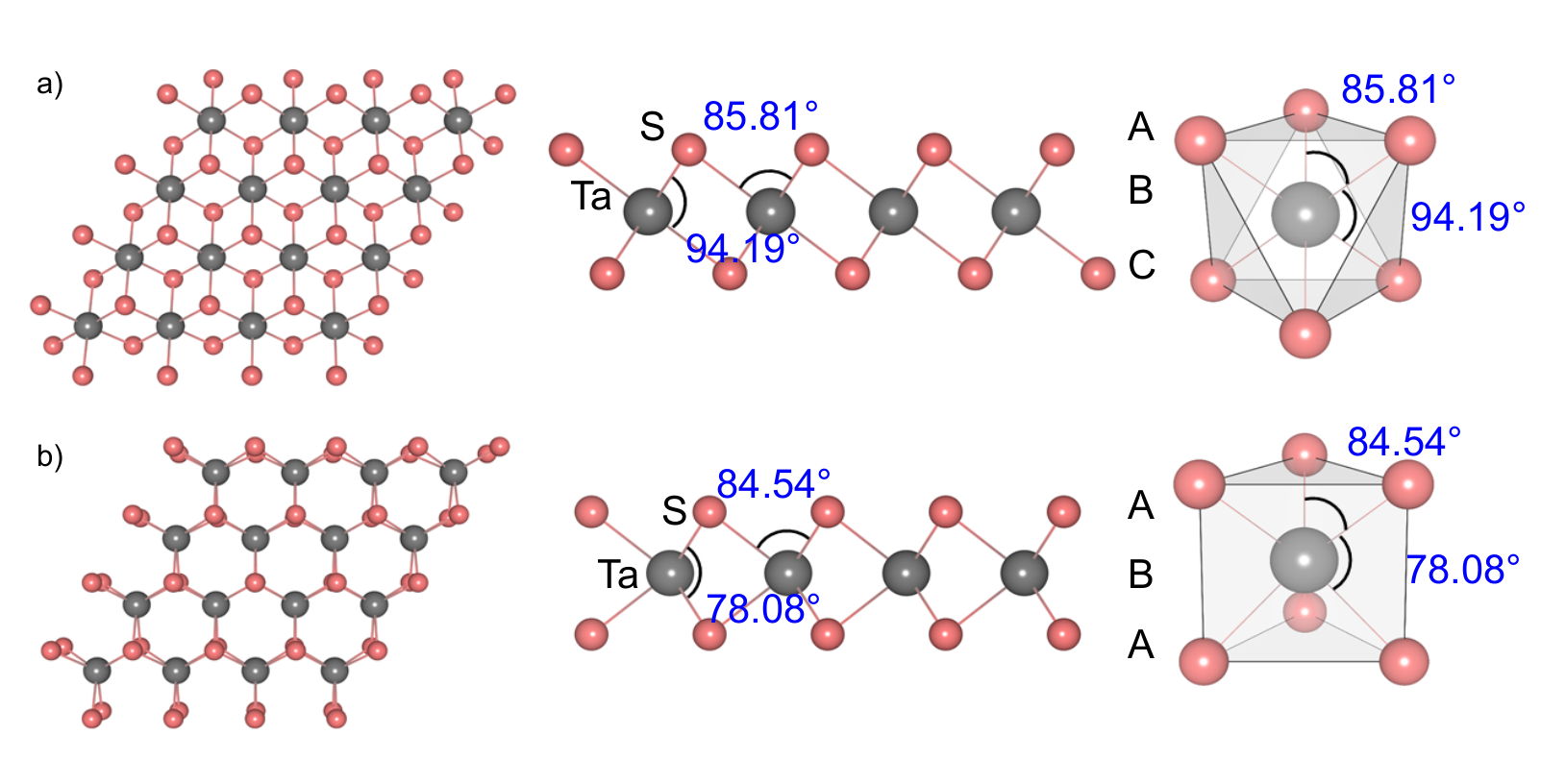}
\caption{\label{fig:structure} Ball-and-stick representation of (a) the $1T$ polymorph, and (b) the $1H$ polymorph of two-dimensional TaS$_2$. The S-Ta-S angles are indicated and the coordination polyhedra for the two phases are displayed in the right panel.}
\end{figure*}

Another degree of freedom one has when constructing Wannier functions is the choice of the set of bands considered.
In Ref.~\cite{scaramucci2015separating}, it was suggested to use this freedom to estimate different contributions to the ligand field splitting of a given set of orbitals. 
For instance, if one is interested in the splitting of the valence $d$ orbitals of a transition metal ion, one can construct Wannier functions by considering only the five bands with predominant $d$ character, provided that those bands constitute an isolated manifold.
In that case, the obtained MLWFs (or PWFs) correspond in general to molecular orbitals with some weight on the ligands due to hybridization. 
Therefore, the calculated splitting contains contributions both from the electrostatic interaction with the negatively charged ligand ions (crystal field), and from hybridization with various ligands' orbitals (ligand field). 
The ligand field can be read on the on-site part of the tight-binding Hamiltonian, i.e. on the diagonal of the matrix  $H_{\alpha\alpha'}^{R=R'}$. 
For a meaningful interpretation of the differences of on-site energies as the ligand field, it is necessary that the matrix $H_{\alpha\alpha'}^{R=R'}$ is diagonal (or at least nearly-diagonal). 
In the following, this will be achieved in two ways.
We will ensure that $H_{\alpha\alpha'}^{R=R'}$ is nearly diagonal by choosing appropriately the direction of the quantization axis $z$.
Small off-diagonal terms, due to the deviation from perfect octahedral symmetry of $1T$ TMDs as well as due to the spin-orbit coupling in both polymorphs, lead to further splittings that we calculate by diagonalizing the matrix.

On the other hand, if a sufficiently large number of bands is included, the $d$-like Wannier functions do not contain tails on the ligands and are atomic-like, so that the extracted splitting can be interpreted as the bare crystal field. 
In order to separate different contributions to the ligand field, one can consider intermediate models by including in the wannierization procedure a set of ligand-derived bands with a certain orbital character, say $p$ character, in addition to the $d$-like bands.
In that case, the $d$-like Wannier functions do not contain any $p$-like tails on the ligands, but could contains tails coming from hybridization with other states. 
The obtained splitting contains therefore no contribution from hybridization with the $p$-states.

A further complication can arise if the bands of interest are entangled with another manifold. 
This complication arises, for instance, in late-group TMDs where the $d$-like and $p$-like manifolds overlap in energy.
In that case, in order to obtain a Wannier Hamiltonian for the desired bands, we perform the disentanglement procedure of Souza, Marzari and Vanderbilt \cite{souza2001}.
In order to derive $n$ Wannier functions from $m > n$ bands in a certain energy window, one needs a criterion to extract an optimal subspace at each $k$-point of the discretized Brillouin zone. 
A possible prescription consists in using orthogonalized projections on a set of trial localized functions with desired orbital character. 
This corresponds to a choice of gauge-fixing matrix defined as in Eq.~\ref{eq:projections}, except that the matrix is rectangular.
Another choice consists in refining the subspace selection via projection by imposing optimal smoothness of the Hilbert space, through the minimization of the gauge-invariant part of the spread functional :
\begin{equation}
\Omega_{I} = \sum_{\alpha} \left( \matrixel{\mathcal{W}_{0\alpha}}{r^2}{\mathcal{W}_{0\alpha}} -\sum_{R\alpha'}| \matrixel{\mathcal{W}_{R\alpha'}}{r}{\mathcal{W}_{0\alpha}}|^2 \right) \, .
\end{equation}
In the following, we shall adopt the optimal smoothness prescription whenever disentanglement is required.
\subsection{Computational details}
Density functional calculations are performed using the \textsc{Quantum ESPRESSO} package \cite{giannozzi2009quantum}. 
The exchange-correlation functional is approximated by the generalized gradient approximation of Perdew, Burke, and Erzernhof (PBE) \cite{perdew_generalized_1996}.
Optimized norm-conserving Vanderbilt pseudopotentials \cite{hamann2013, scherpelz2016implementation, schlipf2015optimization}, from the SG15 library \cite{hamann2013, scherpelz2016implementation, oncv}, are used to described the interaction between valence and core electrons. 
The transition metals' $s$ and $p$ semi-core states are explicitly treated as valence electrons, as well as $f$ states in the case hafnium. 
A plane-wave cutoff of $100$~Ry is used for all the materials considered. 
For tantalum disulfide, we have also used ultrasoft pseudopotentials from the pslibrary \cite{dal_corso_pseudopotentials_2014, uspppage} for the calculation of the projected density of states.
We have checked that the band structures calculated with the two sets of pseudopotentials are identical.
Brillouin zone integration is carried out using a mesh of $24\times24$ $k$-points and a Marzari-Vanderbilt smearing \cite{marzari_thermal_1999} of $10$~mRy.
The structure of each material is obtained by fully relaxing the lattice constant and atomic positions until all the Hellman-Feynman forces are smaller than $10^{-4}$~Ry/Bohr and the pressure is smaller than $0.1$~Kbar. 
About $13$~\AA\ of vacuum is inserted between periodic replicas to simulate a monolayer. 
Wannierization is carried out on a grid of $12\times12$ $k$-points using the Wannier90 code \cite{mostofi_updated_2014}.
\section{The case of TaS$_2$}\label{sec:tas2}
\subsection{spd, pd, and d models}
\begin{figure}
\includegraphics[width=8.5cm]{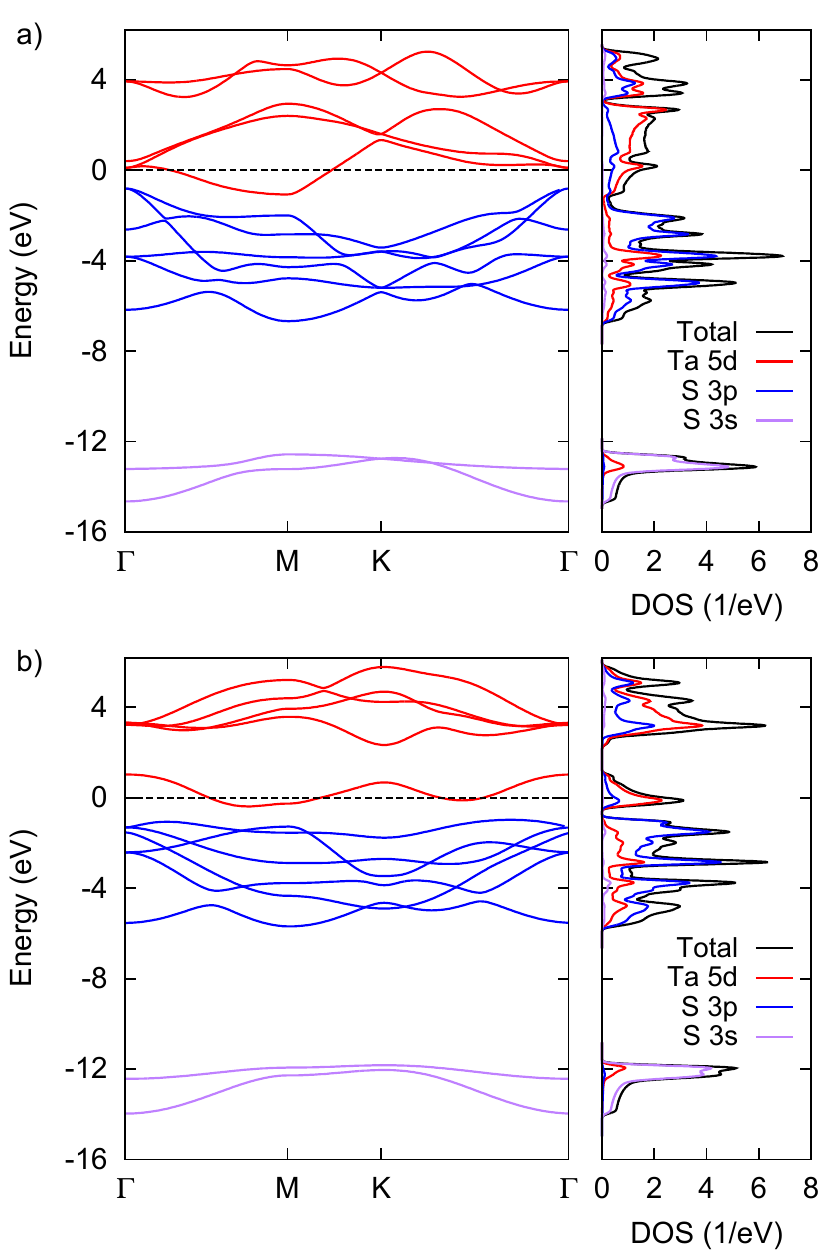}
\caption{\label{fig:pdos} Band structure along high-symmetry directions and projected density of states for (a) $1T$-TaS$_2$, and (b) $1H$-TaS$_2$. The $d$-like bands are shown in red, the $p$-like bands in blue, and the $s$-like bands in purple. The dashed line corresponds to the Fermi level, set to zero.}
\end{figure}
\begin{figure*}[t]
\centering
\includegraphics[width=15cm]{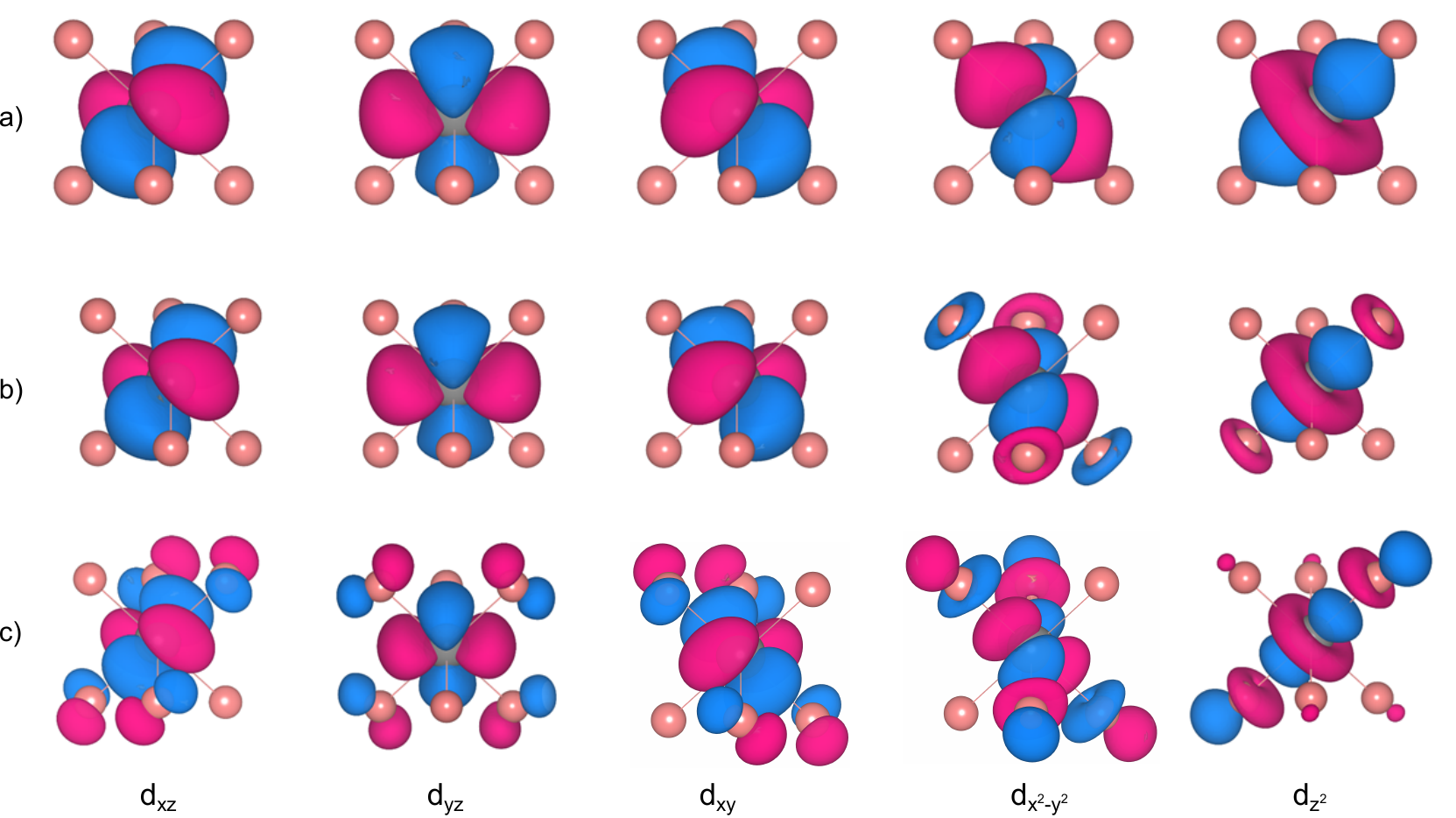}
\caption{\label{fig:wannier1t} Isovalue plots of the five $d$ Wannier functions of $1T$-TaS$_2$ in the (a) $spd$ model, (b) $pd$, and (c) $d$ models.}
\end{figure*}
\begin{figure*}[t]
\centering
\includegraphics[width=15cm]{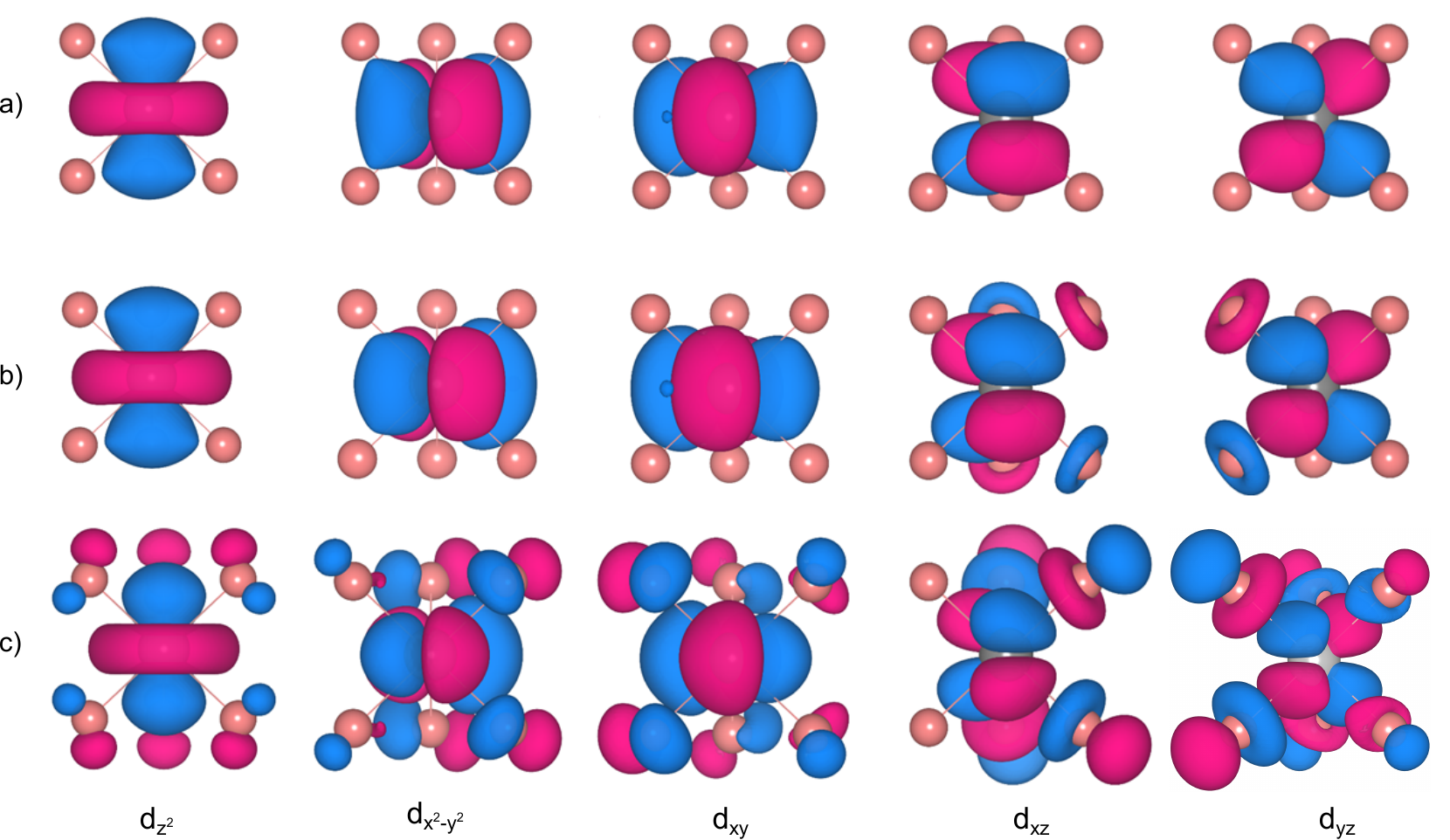}
\caption{\label{fig:wannier1h} Isovalue plots of the five $d$ Wannier functions of $1H$-TaS$_2$ in the (a) $spd$ model, (b) $pd$, and (c) $d$ models.}
\end{figure*}
As an example, we focus our attention on tantalum disulfide (TaS$_2$), a material existing in both polymorphs (in bulk and monolayer forms \cite{wilson1969transition, sakabe2017direct, sanders2016}) and well known for its exotic phase diagram in the $1T$ case, that includes several charge density wave (CDW) transitions and a Mott (or possibly Anderson) insulating phase \cite{di_salvo_low_1977, sipos_mott_2008, law_1t-tas2_2017}.
TaS$_2$ is a convenient case to study because the five $d$-like bands are separated in energy from the lower energy sulfur $p$-like bands, so that there is no need in disentangling the two manifolds.
Moreover, this will allow us to compare our results with the early estimate of Mattheiss \cite{mattheiss1973} in Sec.~\ref{sec:previous}. 

TaS$_2$ belongs to the group V TMDs, so that the formal electronic configuration of Ta$^{4+}$ is $5d^16s^0$.
The crystal structures of the $1T$ and $1H$ phases are shown in Fig.~\ref{fig:structure}.
The unit cell in the two phases contains one formula unit.
As one can see in Fig.~\ref{fig:structure}, the $1T$ and $1H$ phases are characterized by ABC and ABA stackings of the S-Ta-S atomic planes, leading to distorted octahedral (or trigonal antiprismatic) and trigonal prismatic coordinations.
The calculated lattice constants $a_{1T/1H}$,  tantalum-sulfur distances $d_{1T/1H}$, and S-Ta-S angles are summarized in Table~\ref{tab:structure}.
From the calculated S-Ta-S angles $\theta_{1T} = 94.19^{\circ} (85.81^{\circ})$ and $\theta_{1H} = 78.08^{\circ}$ ($84.54^{\circ}$), as shown in Fig.~\ref{fig:structure}, we notice small deviations from a perfect octahedron ($\theta = 90^{\circ}$) and a perfect trigonal prism with equal edges ($\theta \approx 80.8^{\circ}$), respectively.
In Fig.~\ref{fig:pdos}, we show the calculated PBE band structures and density of states.
We have highlighted in color the sulfur $s$- and $p$-like bands, as well as the tantalum $d$-like bands.
As the projected density of state plot shows, there is strong hybridization, especially between the Ta-$d$ and S-$p$ states, indicating the covalent nature of the Ta-S bond.
Nevertheless, throughout this work, we shall continue referring to the five bands shown in red in Fig.~\ref{fig:pdos} as the $d$ bands, to the six bands shown in blue as the $p$ bands, and to the two bands  shown in purple as the $s$ bands. 
We also note that the purple $s$ bands contain non-negligible $d$ character, suggesting that hybridization between $s$ and $d$ states also contributes to the ligand field splitting of the $d$-like states.
\begin{table}
\caption{\label{tab:structure} Calculated structural parameters for the undistorted $1T$ and $1H$ phases of TaS$_2$.}
\begin{ruledtabular}
\begin{tabular}{llll}
 & $a$ (\AA) & $d$ (\AA) & $\theta$  ($^{\circ}$)\\
\hline\\
$1T$ & 3.38 & 2.48 & 94.19\\
$1H$ & 3.34 & 2.48 & 78.08\\
\end{tabular}
\end{ruledtabular}
\end{table}

The band structure suggests that three natural models can be considered to describe valence electrons, that is a 13-band $spd$ model, an 11-band $pd$ model, and a 5-band $d$ model. 
Let us first consider the 13-band $spd$ model. 
We construct Wannier functions, as well as the corresponding Wannier Hamiltonian, as described in Sec.~\ref{sec:methodology}, by including simultaneously the $s$, $p$ and $d$ bands.
For $1T$-TaS$_2$ the two high-energy $d$ bands are slightly entangled with higher-energy bands, not shown in Fig.~\ref{fig:pdos}. 
Therefore, we perform the disentanglement procedure sketched in Sec.~\ref{sec:methodology}. 
A comparison between the disentangled bands and the PBE bands is provided in the Supplemental Information \cite{suppl}.
We obtain two $s$-like and six $p$-like WFs, centered on the sulfur atoms, as well as five $d$-like WFs centered on the tantalum atom. 
In Fig.~\ref{fig:wannier1t}(a) and Fig.~\ref{fig:wannier1h}(a), we present isovalue plots of the obtained $d$-like WFs. 
As the reader will notice, we have chosen different coordinate systems for the two polymorphs, for reasons that we will explain below.
For the $1T$ case, the $z$-axis is defined along one of the Ta-S bonds. 
Since the octahedral symmetry is broken and the S-Ta-S angles are not $90^{\circ}$ (but either $94.19^{\circ}$ or $85.81^{\circ}$), it is not possible to chose at the same time the $x$ and $y$ axes to be exactly parallel to Ta-S bonds. 
On the other hand, for the $1H$ polymorph, the $z$-axis is pointing in the out-of plane direction, while the $x$-axis is chosen parallel to one of the lattice primitive vectors.
\begin{figure}
\centering
\includegraphics[width=8.5cm]{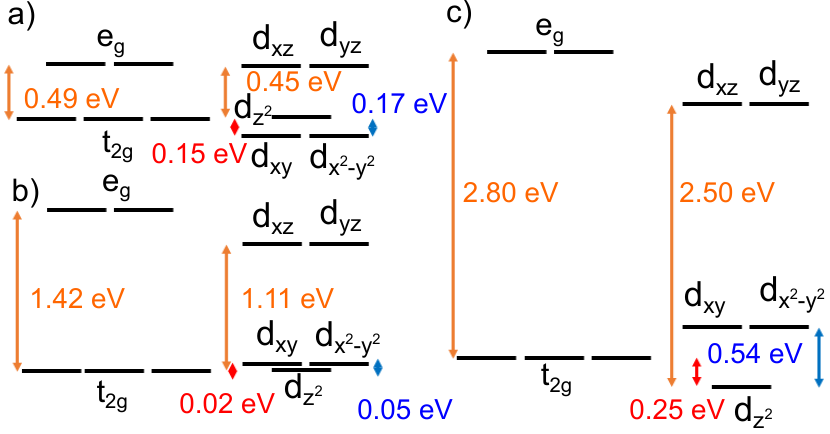}
\caption{\label{fig:crystal_1t1h} Aligned energy levels for $1T$- (left) and $1H$- (right) TaS$_2$ in the (a) 13-band $spd$ model, (b) 11-band $pd$, and (c) 5-band $d$ models. The $t_{2g}$ states in the $1T$ case are the $d_{xy}$, $d_{xz}$, $d_{yz}$ ones, as shown in Fig.~\ref{fig:wannier1t}. The orange arrows correspond to the $t_{2g}$-$e_g$ ($1T$) and $d_{z^2}$-$d_{xz}$ splittings ($1H$). The blue arrows correspond to the splitting of the low-energy triplet in the $1H$ case, and the red arrows indicate the alignment between the lowest-energy levels of the two polymorphs.}
\end{figure}
\begin{table*}
\caption{\label{tab:crystalfield} Summary of spreads ($\Omega$), energy levels ($\epsilon$), and splittings ($\Delta$) for the $d$ WFs in various models. The meaning of the different models is explained in the text. The energy reference is arbitrary, but consistent between different models of a polymorph.}
\begin{ruledtabular}
\begin{tabular}{lllllll}
Model $1T$ & $\Omega_{z^2}$ (\AA$^2$) & $\Omega_{x^2-y^2}$ (\AA$^2$) & $\Omega_{t_{2g}}$ (\AA$^2$) & $\epsilon_{z^2}$ (eV) & $\Delta_{z^2-x^2y^2}$ (eV) & $\Delta_{e_g-t_{2g}}$ (eV) \\
\hline\\
$d$ (5) & 6.36 & 6.33 & 4.68 & 5.44 & 0.03 & 2.80 \\
$pd$ (11) & 2.96 & 2.95 & 2.03 & 2.88 & 0.02 & 1.49   \\
$spd$ (13) & 2.65 & 2.64 & 2.03 & 1.95 & 0.02 & 0.57 \\
$spspd$ (17) & 2.65 & 2.64 & 2.03 & 1.86 & 0.02 & 0.59  \\
$spdds$ (24) & 1.32 & 1.31 & 1.37 & 3.85 & 0.03 & 0.88  \\
$spddsp$ (27) & 1.20 & 1.20 & 1.42 & 3.88 & 0.03 & 0.92  \\
\hline\\
Model $1H$ & $\Omega_{z^2}$ (\AA$^2$) & $\Omega_{xy}$ (\AA$^2$) & $\Omega_{xz}$ (\AA$^2$) & $\epsilon_{z^2}$ (eV) & $\Delta_{xy-z^2}$ (eV) & $\Delta_{xz-z^2}$ (eV) \\
\hline\\
$d_{z^2}$/$d_{xy,x^2-y2,xz, yz}$ (1/4) & 15.59 & 9.73 & 5.34 & 0.84 & 3.27 & 4.29\\
$d$ (5) & 3.93 & 4.8 & 5.34 & 2.66 & 0.54 & 2.47 \\
$pd$ (11) & 2.13 & 2.37 & 2.39 & 1.65 &  0.05 & 1.11 \\
$spd$ (13) & 2.13 & 2.27 & 2.21 & 1.65 & -0.17 & 0.45 \\
$spspd$ (17) & 2.13 & 2.28  & 2.21 & 1.62 & -0.19 & 0.47 \\
$spdds$ (24) & 1.51 & 1.26 & 1.24 & 3.18 & -0.09 & 0.65 \\
$spddsp$ (27) & 1.15 & 1.21 & 1.22 & 3.32 & -0.07 & 0.52 \\
\end{tabular}
\end{ruledtabular}
\end{table*}
As one can see in Figs. \ref{fig:wannier1t} and \ref{fig:wannier1h}, the $d$-like WFs in the $spd$ model are atomic-like and do not contain any visible hybridization with the sulfur $3s$ and $3p$ orbitals. 
Therefore, the calculated energy differences between the on-site energies of the WFs, obtained by inspecting the $d$ block of the Wannier Hamiltonian, should be a good approximation to the bare crystal field, coming from the electrostatic interaction with the negatively charged sulfur ions.
The calculated energy splittings are reported schematically in Fig.~\ref{fig:crystal_1t1h}(a). 

For $1T$-TaS$_2$, we obtain a three-below-two energy splitting pattern, as expected from crystal field theory.
The three $d_{xz}$, $d_{yz}$ and $d_{xy}$ WFs have on-site energies that are degenerate within $0.01$~eV. 
On the other hand, the two $d_{x^2-y^2}$  and $d_{z^2}$ WFs are higher in energy by $0.57$~eV, with a small difference of $ \Delta_{z^2-x^2y^2}^{(13)}=\epsilon_{z^2}^{(13)}-\epsilon_{x^2-y^2}^{(13)} = 0.02$~eV, where $\epsilon_{\alpha}^{(13)}$ refers to the on-site energy of the WF of type $\alpha$ in the 13-band model.
In the following, we shall refer to the $d_{xz}$, $d_{yz}$ and $d_{xy}$ WFs as the $t_{2g}$ triplet, and to the $d_{x^2-y^2}$  and $d_{z^2}$ WFs as the $e_g$ doublet, because the octahedral symmetry remains a useful approximate symmetry. 
Whenever discussing the crystal or ligand field splitting between the $e_g$ doublet and the $t_{2g}$ triplet, we actually mean the difference between the average on-site energies.

For $1H$-TaS$_2$, because of the trigonal prismatic coordination, crystal field theory predicts a splitting of the $d$ levels into a singlet $a_1'$ (following the notation of Ref.~\cite{huisman1971trigonal}), a low-energy doublet $e'$ and a high-energy doublet $e''$.
As shown in Fig.~\ref{fig:crystal_1t1h}a), we have obtained, in the $spd$ model, two degenerate doublets $d_{xy}/d_{x^2-y^2}$ and $d_{xz}/d_{yz}$, corresponding to the $e'$ and $e''$ doublets, respectively.
The on-site energies within a doublet differ by less than $0.005$~eV.
Contrary to the $1T$ case where the lowering of the symmetry from octahedral to trigonal antiprismatic leads to intrinsic lifting of degeneracies, we interpret those small differences being due to the wannierization procedure that does not preserve exactly the symmetries.
In the $spd$ model, we find that the $d_{z^2}$ $a_1'$ singlet is $0.17$~eV higher in energy compared to the $d_{xy}/d_{x^2-y^2}$ doublet, and $0.45$~eV lower than the $d_{xz}/d_{yz}$ doublet.
In Ref.~\cite{huisman1971trigonal}, Huisman \textit{et al.} considered a point-charge model to calculate the crystal field splitting in a trigonal prismatic coordination.
Huisman \textit{et al.} obtained that the relative positions of the singlet and low-energy doublet depends sensitively both on the angle between the ligand and the $z$-axis, and on the spread of the orbitals. 
Using parameters assumed relevant for MoS$_2$, it was estimated that the singlet $a_1'$ should be higher in energy than the $e'$ doublet.
As we shall discuss in Sec.~\ref{sec:trend} where we study other TMDs, we have consistently found that, in the $spd$ model, the $d_{z^2}$ singlet is slightly higher in energy than the $d_{xy}/d_{x^2-y^2}$ doublet, except for $d^6$ TMDs where the whole crystal field splitting is reversed.

Both for the $1T$ and $1H$ polymorphs, the calculated energy splittings are rather small ($\sim 0.5$~eV) compared to the overall bandwidth of the five $d$ bands ($\sim 7$~eV). 
This is in agreement with the intuitive expectation that for covalent bonding, electrostatic effects should not be dominant, while the hybridization with the ligands' valence orbitals is important.
We have then considered an $11$-band $pd$ model, by constructing six $p$ WFs and five $d$ WFs from the bands shown in blue and red in Fig.~\ref{fig:pdos} simultaneously.
In Figs.~\ref{fig:wannier1t}(b) and \ref{fig:wannier1h}(b), we present plots of the corresponding $d$ WFs. 
We see that, although the isovalue for the plots is the same as for the WFs in the $spd$ model, certain WFs exhibit considerable weight on the sulfur atoms, typical of molecular orbitals with antibonding character.
Indeed, the on-site energies of some of the $d$ WFs are shifted upward in energy compared to the $spd$ model, as we summarize in Table~\ref{tab:crystalfield}.
The differences of on-site energies between the $spd$ and $pd$ models can be interpreted as the hybridization energy between $s$ and $d$ orbitals \cite{scaramucci2015separating}.

For the $1T$ case, we obtain hybridization energies $\epsilon^{(11)}_{e_g}-\epsilon^{(13)}_{e_g}= 0.93$~eV, and $\epsilon^{(11)}_{t_{2g}}-\epsilon^{(13)}_{t_{2g}}=0$~eV.
Also, the spread $\Omega$ of the $e_g$ WFs, defined in Sec.\ref{sec:methodology}, increases from $2.65$~\AA$^2$  to $2.95$~\AA$^2$, whereas the spread of the $t_{2g}$ WFs is identical in the two models. 
This indicates lack of hybridization between $t_{2g}$ $d$ orbitals and sulfur $s$ orbitals, consistent with the observation in the projected density of states plot, in Fig.~\ref{fig:pdos}(a), that the three low-energy $d$ bands have negligible sulfur $s$ character.
Therefore, we conclude that the hybridization with the $s$ orbitals leads to a significant increase of the ligand field splitting of $0.93$~eV, as represented in Fig.~\ref{fig:crystal_1t1h}(b). 
It is worth mentioning that the on-site energies of the $p$ WFs change only slightly between the two models, as one would expect by noticing in Fig.\ref{fig:pdos}(a) that the $s$ bands have a negligible $p$ character.
We obtain a small difference between the average of the on-site energies of $\bar{\epsilon}_p^{(11)}-\bar{\epsilon}_p^{(13)} = 0.06$~eV.

For the $1H$ polymorph, as we report in Table~\ref{tab:crystalfield}, $sd$ hybridization leads to an increase of the on-site energy of two two doublets, while the $d_{z^2}$ singlet remains unaffected. 
The splitting $\Delta_{xz-z^2}= \epsilon_{xz}-\epsilon_{z^2}$ therefore increases from $0.45$~eV to $1.11$~eV in the $pd$ model, meaning a hybridization energy of $0.66$~eV for the high energy $d_{xz}/d_{yz}$ doublet. 
As shown in Fig.~\ref{fig:crystal_1t1h}, the splitting $\Delta_{xy-z^2}= \epsilon_{xy}-\epsilon_{z^2}$ has a different sign compared to the $spd$ model, because of the $sd$ hybridization energy  $\epsilon^{(11)}_{xy}-\epsilon^{(13)}_{xy}= 0.22$~eV for the low-energy $d_{xy}/d_{x^2-y^2}$ doublet.

In order to account for the $pd$-hybridization contribution to the ligand field, we consider a 5-band $d$-only model, constructed by including only the five $d$ bands during the wannierization procedure.
In Fig.~\ref{fig:wannier1t}(c), we show isovalue plots of the derived WFs. 
We see that all five WFs have large weight on the sulfur atoms and resemble molecular orbitals with large $pd$ antibonding hybridization.
We observe that $\pi$-bonding occurs for the $t_{2g}$ WFs, while $\sigma$-bonding takes place in the case of the $e_g$ WFs.
This leads to a larger hybridization energy for the $e_g$ WFs, $\epsilon_{z^2}^{(5)}-\epsilon_{z^2}^{(11)}= 2.56$~eV, causing an increase of the $e_g$-$t_{2g}$ splitting of $1.31$~eV, so that it is $2.80$~eV in the 5-band model.
\begin{figure}
\includegraphics[width=8.5cm]{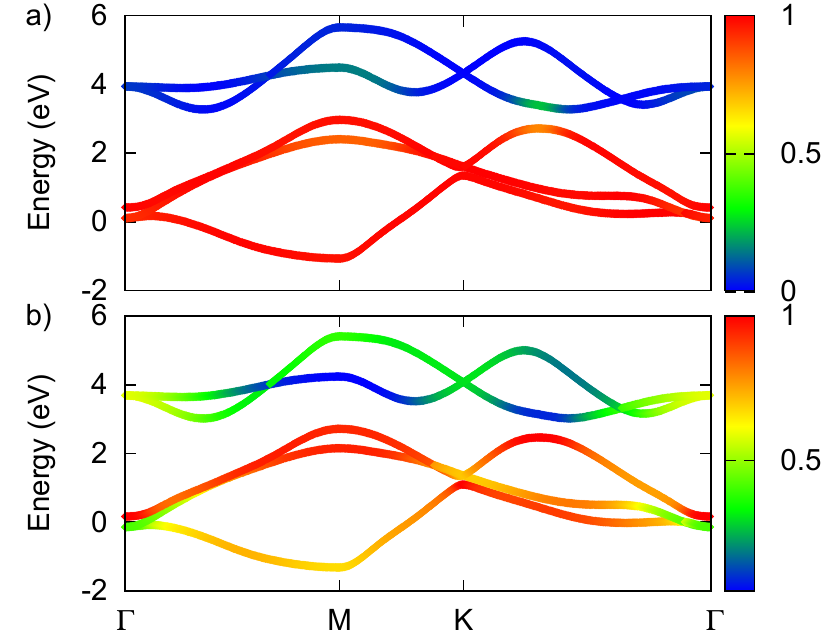}
\caption{\label{fig:1t_5band} Band structure for the 5-band model of $1T$-TaS$_2$ with orbital weight as a color code of the (a) $t_{2g}$ ($d_{xy}$, $d_{xz}$ and $d_{yz}$, with $z$ along a Ta-S bond) WFs, and (b) $d_{z^2}$, $d_{xy}$ and $d_{x^2-y^2}$ WFs, with $z$ in the out-of-plane direction. The Fermi level is set to zero.}
\end{figure}
\begin{figure}
\includegraphics[width=8.5cm]{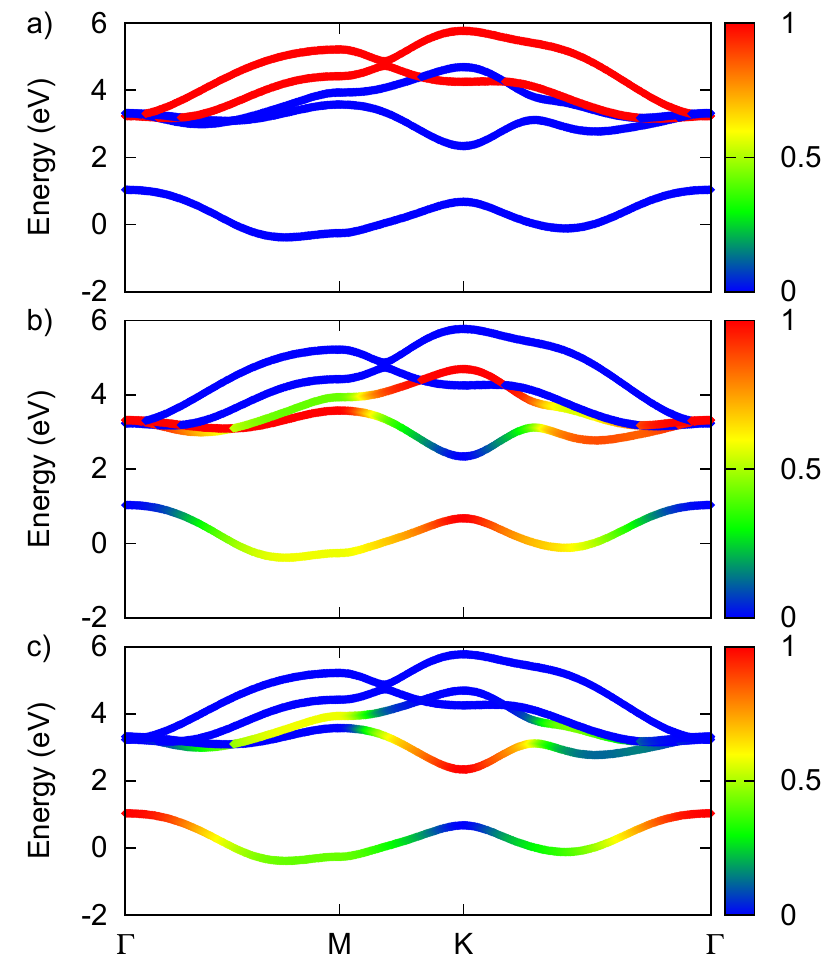}
\caption{\label{fig:1h_5band} Band structure for the 5-band model of $1H$-TaS$_2$ with orbital weight as a color code of the (a) two $d_{xz}$ and $d_{yz}$ WFs, (b) two $d_{xy}$ and $d_{x^2-y^2}$ WFs, and (c) $d_{z^2}$ WF. The Fermi level is set to zero.}
\end{figure}

In Fig.~\ref{fig:wannier1h}(c), we show isovalue plots of the five WFs in the $d$-only model for $1H$-TaS$_{2}$.
As for the $1T$ polymorph, it is apparent that all five WFs resemble molecular orbitals with antibonding character. 
The $d_{z^2}$ WF forms $\pi$-antibonding combinations with the sulfur $p$ orbitals.
The calculated $pd$ hybridization energy is given by $\epsilon_{z^2}^{(5)}-\epsilon_{z^2}^{(11)}= 1.01$~eV. 
The $d_{xy}/d_{x^2-y^2}$ and $d_{xz}/d_{yz}$ doublets interact more strongly with the ligands' $p$ orbitals, leading to increased ligand field parameters $\Delta_{xy-z^2}^{(5)}= 0.54$~eV and $\Delta_{xz-z^2}= 2.5$~eV, as shown in Fig.~\ref{fig:crystal_1t1h}(c).

It is now appropriate to discuss in more detail the different choices of coordinate system for the $1T$ and $1H$ polymorphs.
In the literature, crystal field arguments are often given to qualitatively describe the electronic structure of dichalcogenides.
For $1T$ TMDs, while the ligand field spitting is often discussed using the same coordinate system as here (see e.g. Refs.~\cite{chhowalla_chemistry_2013, whangbo1992analogies, chentrimer2018}), some authors discuss the low energy triplet in terms of $d_{z^2}$, $d_{xy}$ and $d_{x^2-y^2}$ orbitals with the $z$ axis pointing in the out-of-plane direction, following Mattheiss \cite{mattheiss1973}.  
In Fig.~\ref{fig:1t_5band}(a), we show the band structure of $1T$-TaS$_2$ with the orbital weight of the three $t_{2g}$ WFs in the 5-band model that we have discussed above.
We see that the $t_{2g}$ WFs give rise to the three low-energy bands, and that the hybridization with the $e_g$ WFs is very weak. 
We interpret the weak $t_{2g}$-$e_g$ hybridization as resulting from weak the distortion of the octahedral symmetry.
In Fig.~\ref{fig:1t_5band}(b), we show the orbital weight of the three low-energy orbitals with the $z$-axis defined out-of plane, as for the $1H$ polymorph.
With this choice of axes, the on-site part of the Wannier Hamiltonian contains large off-diagonal matrix elements $H^{R=R'}_{\alpha\neq\alpha'}$ ($\sim 1.2$~eV in the $d$ model, against $\sim 0.1$~eV with the other coordinate choice), so that the three low-energy WFs strongly hybridize with the two high-energy WFs, as can be seen in Fig.~\ref{fig:wannier1t}(b).
By inspection of the on-site energies of the WFs, we obtain the same ordering as in the $1H$ case, with a splitting $\epsilon_{xy}^{(5)}-\epsilon_{z^2}^{(5)}=0.74$~eV. 
However, for both choices of coordinate systems, the same three-below-two splitting pattern is obtained upon diagonalizing the on-site Hamiltonian matrix, with identical energy eigenvalues.
Therefore, our choice of coordinate system is motivated by the fact that the corresponding splitting of on-site energies leads to a better approximate picture for the electronic structure.
On the other hand, for the $1H$ polymorph, the on-site Hamiltonian is exactly diagonal when the $z$-axis is chosen out-of-plane. 
From Fig.~\ref{fig:1h_5band}(a), it is evident that the two high-energy $d_{xz}/d_{yz}$ WFs are decoupled from the three low-energy WFs.

It is worth mentioning that, although we have defined the WFs by projection and have not performed the localization procedure, the obtained PWFs are very close to maximal localization, with a nearly zero imaginary part.
In the 5-band model, the ligand field spittings calculated with PWFs and MLWFs are nearly identical. 
The main difference arises in the 13-band model, where the localization procedure admixes the $s$ WFs with other WFs, leading to a slightly reduced total spread but to less localized $d$-like WFs. 
We have also found that, in certain cases, the localization procedure leads to a change in coordinate system.
Therefore, we have adopted PWFs instead of MLWFs, giving us a better control of the orbital character and the coordinate system.
\subsection{Semi-core and high-energy states}
In the 13-band model, the $d$-like WFs are atomic-like. 
However, it is expected that they are even more localized for models derived from a larger number of bands. 
We have therefore first considered including tantalum $5s$ and $5p$ semi-core states in the construction of the WFs. 
The changes in the spread of the WFs, on-site energies and splittings are summarized in Table~\ref{tab:crystalfield}.
As expected, the effect of the inclusion of the semi-core states on the calculated crystal field splitting and on the spread of the $d$ WFs is very weak, as core electrons are non-bonding by definition. 

When plotting the $d$ WFs of the 13-band model with a sufficiently small isovalue, one can recognize tails on the sulfur atoms that resemble $d$ electrons. 
This means that $d$ WFs are in fact bonding combinations of Ta $5d$ and S $3d$ orbitals. 
Since the excited-state bands above the Ta $d$-bands are highly entangled, it is not possible to isolate a set of bands corresponding to the sulfur $d$ electrons. 
In order to assess the effect of Ta-$d$/S-$d$ hybridization on the crystal field, we include $40$ excited states above the Ta $d$ bands and disentangle the Ta $6s$ and S $3d$ bands, keeping the 13 valence bands frozen.
The tantalum $6s$ states are explicitly kept because they are lower in energy than the sulfur $d$ ones.
Therefore, we obtain a 24-band $spdds$ model, describing a finite set of excited states in addition to the valence bands.
The corresponding band structure for $1T$-TaS$_2$ is shown in the Supplementary Information \cite{suppl}. 
The $d$ WFs are more localized and have a higher energy than in the 13-band model, as summarized in Table~\ref{tab:crystalfield}. 
The energy splittings are somewhat increased compared to the 13-band model ($0.88$~eV against $0.57$~eV for $1T$-TaS$_2$), indicating a small \emph{negative} contribution to the total ligand field.
We have also considered a 27-band $spddsp$ model, including Ta $6p$ states, that yields similar results.

To summarize, we conclude that the total $e_g$-$t_{2g}$ splitting of $2.80$~eV in $1T$-TaS$_2$ is the result of positive contributions from electrostatic effects and hybridization with ligands' $3s$ and $3p$ states, all of the order of $\sim 1$~eV, as well as a smaller negative contribution of $~0.3$~eV due to the formation of bonding combinations with the higher energy ligands' $3d$ states.
Also, the $d_{xy}$-$d_{z^2}$ splitting of $0.54$~eV  in $1H$-TaS$_2$ comes from negative contributions due to electrostatic effects and hybridization with $3d$ states (of the order of $\sim 0.1$~eV each), and positive contributions from hybridization with $3s$ and $3p$ states ($0.22$ and $0.49$~eV, respectively).
\subsection{Spin-orbit coupling}\label{sec:soc}
\begin{figure}
\centering
\includegraphics[width=7.5cm]{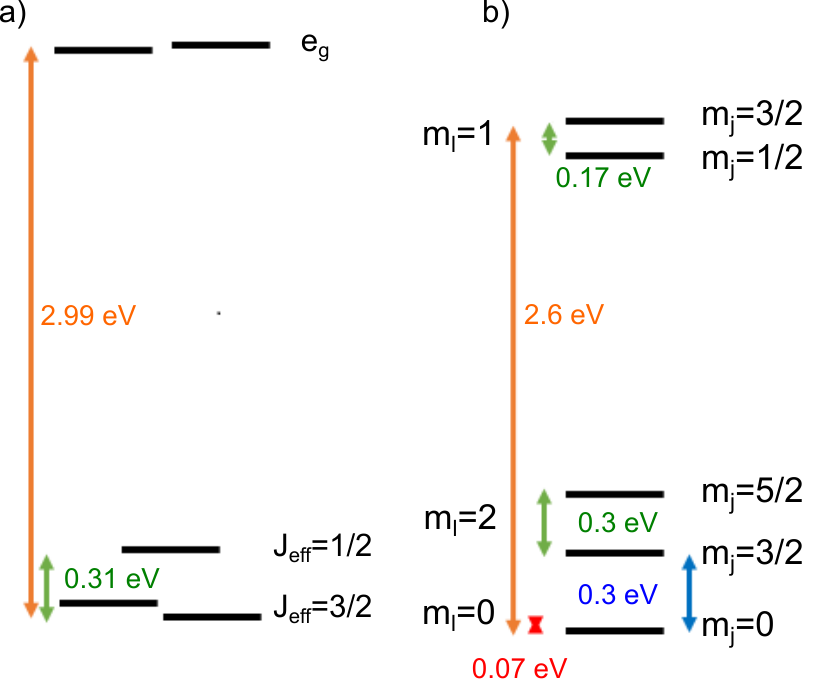}
\caption{\label{fig:soc} Aligned ligand field energy diagram, including the SOC, for (a) $1T$-TaS$_2$, and (b) $1H$-TaS$_2$. The orange arrows indicate the splittings between the lowest- and highest-energy states, the green arrows indicate the splittings of otherwise degenerate states induced by the SOC, the blue arrow indicates the ligand field splitting of the low-energy states in the 1H case, and the red arrow correspond to the alignment between the lowest-energy state of the two polymorphs.}
\end{figure}
The spin-orbit coupling (SOC) is strong in many TMDs and leads to many interesting effects, such as topological phases in distorted $d^2$ $1T$ TMDs \cite{qian2014quantum}, or Ising superconductivity in single-layer $d^1$ and doped $d^2$ $1H$ TMDs \cite{xi_ising_2016, lu_evidence_2015}.
Here, we investigate the effect of the SOC on the calculated ligand field splittings in $1T$- and $1H$-TaS$_2$. 
The SOC introduces off-diagonal imaginary matrix elements in the on-site part of the Hamiltonian, lifting degeneracies.
In Fig.~\ref{fig:soc}, we present the calculated ligand field diagrams in the $d$ model with and without the SOC.

For the $1T$ polymorph, the SOC splits the $t_{2g}$ manifold into a lower-energy $J_{\mathrm{eff}=3/2}$ doublet and a higher-energy  $J_{\mathrm{eff}}=1/2$ singlet. 
The calculated splitting of $0.31$~eV is modest compared to the bandwidth ($\sim 3.5$~eV) of the three $t_{2g}$ bands, as well as compared to the ligand field splitting  $\Delta_{e_g-t_{2g}}^{(5)} = 2.80$~eV.
On the other hand, the $e_g$ doublet remains nearly degenerate (within $0.03$~eV). 
The degeneracy of the $J_{\mathrm{eff}}=3/2$ doublet is lifted by $0.07$~eV.
These findings are in agreement with crystal field theory, that predicts a splitting of the $t_{2g}$ shell induced by the SOC, but not of the $e_g$ shell.

For $1H$-TaS$_2$, Fig.~\ref{fig:soc} shows that the SOC splits the low-energy and high-energy doublets by $0.3$~eV and $0.17$~eV, respectively.
We note that the SOC splitting is of the same magnitude as the ligand field $\Delta_{xy-z^2}$ splitting.
\section{Crystal field and the relative stability of the $1T$ and $1H$ phases}\label{sec:stability}
\begin{figure}
\centering
\includegraphics[width=8.5cm]{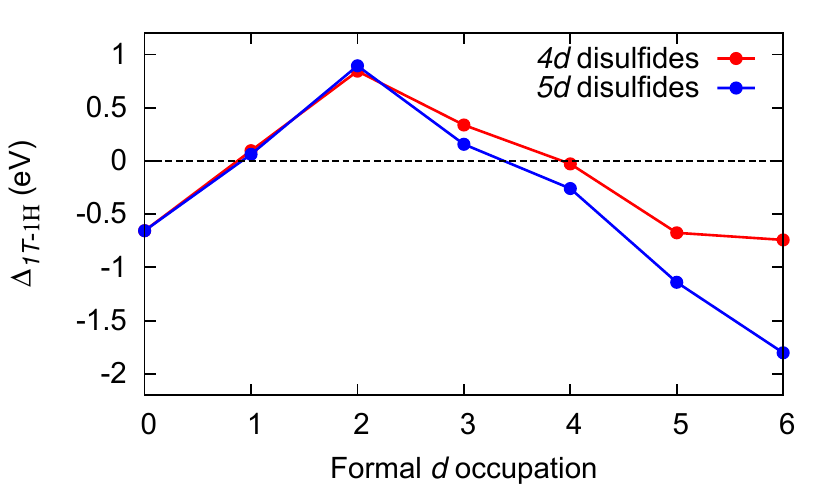}
\caption{\label{fig:relative} Calculated total energy difference, per formula unit, between the $1T$ and $1H$ phases of  $4d$ and $5d$ disulfides, as a function of the formal number of $d$ electrons. A negative energy indicates higher stability of the $1T$ phase.}
\end{figure}
Having estimated crystal field and ligand field parameters for TaS$_2$, we proceed to discussing their relevance in determining the relative stability of the $1T$ and $1H$ phases.
In Fig.~\ref{fig:relative}, we present the calculated total energy difference per formula unit $\Delta_{1T-1H}= (E_{1H}-E_{1T})$ for the series of material MS$_{2}$, with M belonging to the the $4d$ and $5d$ rows of transition metals. 
Fig.~\ref{fig:relative} shows that the $1H$ phase is energetically slightly more favourable than the $1T$ phase in the $d^1$ (NbS$_2$ and TaS$_2$) and $d^3$ cases (TcS$_2$ and ReS$_2$), highly favourable in the $d^2$ case (MoS$_2$ and WS$_2$), and unfavourable for any other filling.
Note that some of the materials calculated in this plot, such as OsS$_2$, do not exist in nature in either the $1T$ or $1H$ polymorph (in neither bulk nor monolayer form).
However, it is instructive to calculate their properties in order to discuss trends. 
In Fig.~\ref{fig:relative}, we notice a discontinuity at the $4d^6$ point, corresponding to PdS$_2$. 
Indeed, the $1H$ polymorph of PdS$_2$ relaxes to a structure with a short S-S distance of $\approx 2$~\AA, leading to a qualitatively different electronic structure. 
Nevertheless, the corresponding structure is still highly unfavourable with respect to the $1T$ phase.
In light of the preceding, we are now in a position to assess the role of crystal and ligand field effects in determining this trend. 
As in the previous section, we shall focus on the case of monolayer TaS$_2$ as an example.
As we will show in the next section by discussing trends across the periodic table, the physics discussed here is not unique to TaS$_2$ but applies to the entire family of TMDs, because of the universality of the band structure.
The discussion also applies to bulk materials as long as interlayer couplings are not too strong. 

\begin{figure}
\centering
\includegraphics[width=8.5cm]{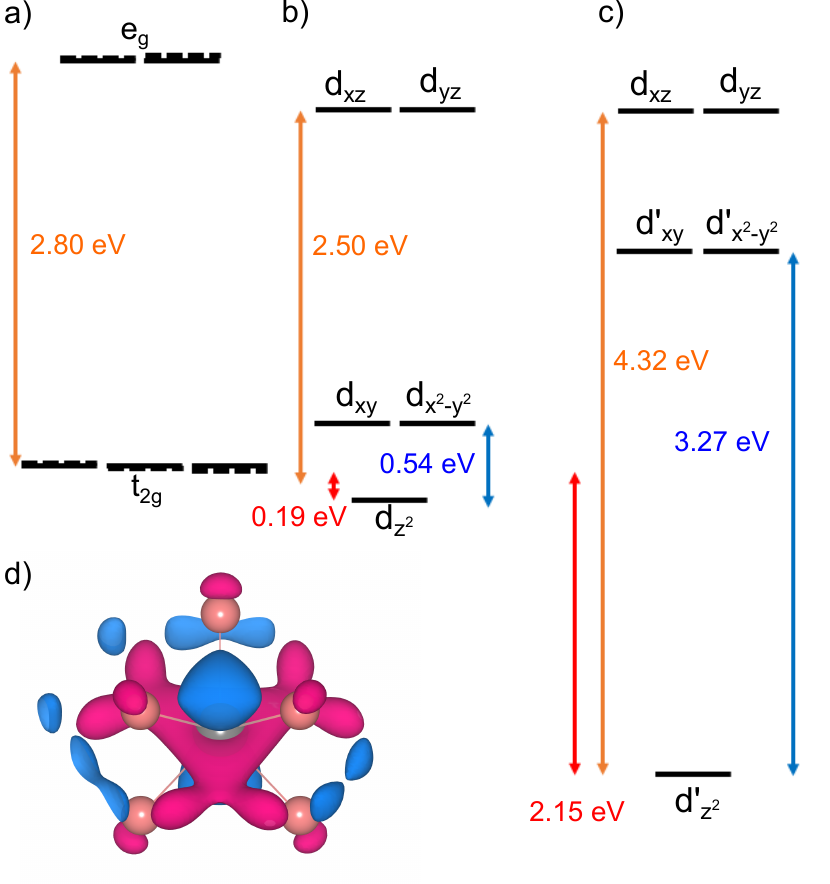}
\caption{\label{fig:crystal_interorbital} (a)-(b) Aligned ligand field diagrams for $1T$ and $1H$ TaS$_2$. The dashed lines correspond to the eigenvalues of the on-site Hamiltonian matrix for the $1T$ case. (c) Energy levels in the 1-band/4-band models (the meaning of which is explained in the text) for $1H$-TaS$_2$. The orange arrows correspond to the $t_{2g}$-$e_g$ ($1T$) and $d_{z^2}$-$d_{xz}$ splittings ($1H$). The blue arrows correspond to the splitting of the low-energy triplet in the $1H$ case, and the red arrows indicate the alignment between the lowest-energy levels of the two polymorphs. (d) Isovalue plot of the $d'_{z^2}$ Wannier function. }
\end{figure}

In Fig.~\ref{fig:crystal_1t1h}, we have plotted the aligned crystal and ligand field energy diagrams for $1T$- and $1H$-TaS$_2$.
In the 13-band model, the $d_{z^2}$ state is actually slightly higher in energy compared to the $d_{x^2-y^2}$ and $d_{xy}$ states, so that the bare crystal field does not favour a $d_{z^2}^1$ configuration for the $1H$ polymorph.
However, as one can infer from the 11-band and 5-band models, hybridization effects with the ligand $s$ an $p$ states pushes the $d_{xy}$ and $d_{x^2-y2^2}$ states $\sim 0.5$~eV higher in energy.
Hence, in a local picture neglecting inter-site hoppings, the ground-state configuration for $n < 2$ $d$ electrons is obtained by partially filling the $d_{z^2}$ state, in agreement with the standard ligand field argument. 
From Fig.~\ref{fig:crystal_1t1h}~(c), we observe that, in the $d$ model, the $d_{z^2}$ level in the $1H$ polymorph is $0.23$~eV below the $t_{2g}$ levels of the $1T$ polymorph. 
Note that the energy levels were aligned by equalizing the vacuum energies. 
Without such alignement, the energy difference is somewhat smaller, i.e. $0.06$~eV.
The calculated stabilizing energy of $0.23$~eV is small compared to the amplitude of the bandwidths, or even compared to lifting of degeneracies induced by the octahedral symmetry breaking and by the spin-orbit coupling.
Indeed, the stablilizing energy of the $d_{z^2}$ singlet is reduced to $0.19$~eV when the lifting of degeneracy of the $t_{2g}$ states is taken into account (see Fig.~\ref{fig:crystal_interorbital}(a), where we have represented with dashed lines the eigenenergies of the on-site Hamiltonian). 
When the SOC is included, as is shown in Fig.~\ref{fig:soc}, the energy gain of the $d_{z^2}$ level compared to the $J_{\mathrm{eff}}=3/2$ doublet is even more reduced to $0.07$~eV (note that the SOC does not improve the relative stability of the $1T$ polymorph of TaS$_2$).
While the ligand field \emph{does} actually favour the $1H$ polymorph for $n < 2 $ and disfavours it for $n>2$ , this effect appears to be rather weak and insufficient to explain the calculated trend presented in Fig.~\ref{fig:relative}.

In Fig.~\ref{fig:1h_5band}, we plot the band structure of the $1H$ phase in the 5-band model with the orbital weights of the three groups of WFs in pannels (a), (b) and (c). 
As we have already discussed in Sec.~\ref{sec:tas2}, the $d_{xz}/d_{yz}$ doublet is perfectly decoupled from the three low-energy Wannier functions. 
This is guaranteed by symmetry since the two groups of orbitals have a different parity under the exact mirror symmetry, i.e. they pick a different sign under the $z \to -z$ transformation. 
On the other hand, it is clear from Fig.~\ref{fig:1h_5band}(b)-(c) that the $d_{z^2}$ WF strongly hybridizes with the $d_{xy}$ and $d_{x^2-y^2}$ WFs, except at the high-symmetry $\Gamma$ and K points. 

From Fig.~\ref{fig:1h_5band}(b) and (c), we see that the isolated low-energy band is not only of $d_z^2$ character, but contains strong weight from the $d_{xy}$ and $d_{x^2-y^2}$ WFs \cite{mattheiss1973, kertesz1984octahedral}. 
The emergence of this isolated band is therefore not directly related to the ligand field splitting $\Delta_{z^2,xy}$, as often believed, but to the hybridization between the three low-energy Wannier functions.
This was first emphasized by Mattheiss based on his early band structure calculations of layered TMDs \cite{mattheiss1973}.
Mattheiss noticed the mixed orbital character of the low-energy band, and showed (in the case of $2H$-MoS$_2$) that the gap closes if the interorbital hoppings are set to zero.
More recently, Isaacs and Marianetti gave a similar argument for $1H$-VS$_2$ \cite{isaacs2016}.
Considering an 11-band $pd$ model derived from MLWFs, they showed that the low-energy isolated band is no longer isolated if the direct $d$-$d$ hoppings are set to zero.

In the following, we argue that this gap opening can be understood from a simple intuitive band structure effect.
At the $\Gamma$ point, the $d_{z^2}$ and $d_{xy}-d_{x^2-y^2}$ bands cannot hybridize because they belong to different representation of the point group.
The gap at the $\Gamma$ point ($\sim 2$~eV) is much larger than the calculated ligand field, as it contains large contributions from band structure effects, especially from nearest-neighbor hoppings (NNHs). 
The $d_{z^2}$ NNHs $t_{z^2z^2}$ are negative and equal in all directions so that the $d_{z^2}$ band at the $\Gamma$ point has an energy given by $\epsilon_{z^2}(k=\Gamma) \approx \epsilon_{z^2}^{(5)}-6|t_{z^2z^2}|$, with $|t_{z^2z^2}|=0.17$~eV. 
On the other hand, $d_{xy}/d_{x^2-y^2}$ WFs have hoppings with different signs along different directions, leading to a partial cancellation of NNH effects on the band energy at the zone center.
The result is a band energy at the $\Gamma$ point higher than the on-site energy by $\sim 0.8$~eV.
Since the $d_{z^2}-d_{z^2}$ hoppings are negative, the $d_{z^2}$ band disperses to higher energy as the momentum moves away from the $\Gamma$ point, while the $d_{xy}-d_{x^2-y^2}$ bands split and disperse to lower energy.
As Fig.~\ref{fig:1h_5band} shows, the crossing between those bands is avoided, resulting in a rather large gap because the NNHs between the $d_{z^2}$ and $d_{xy}/d_{x^2-y^2}$ WFs are large ($\sim 0.6$~eV).
At the K point, hybridization between the two sets of WFs is also prevented by symmetry, so that the corresponding state of the low-energy band is given by the bottom of one of the two $d_{xy}/d_{x^2-y^2}$-bands.
We note that the interorbital hybridization is maximal at the bottom of the low-energy band, suggesting a strong stabilization effect for the $1H$ polymorph when the corresponding states are filled.

In order to estimate the contribution from interorbital hybridization to the stabilization of the $1H$ phase, we consider another model, derived by performing wannierization by considering the lower $d$ band (1-band $d'_{z^2}$ model) and the four higher-energy $d$ bands (4-band model) \emph{separately}.
In Fig.~\ref{fig:crystal_interorbital}(c), we show the $d_{z^2}$-like WF for the 1-band model. 
In the 1-band model, the obtained $d_{z^2}$-like Wannier function is strongly distorted compared to the 5-band model, while still resembling a $d_{z^2}$ orbital.
In the following, we shall refer to it as the $d'_{z^2}$ WF, and to the two higher-energy WFs as $d'_{xy}$ and $d'_{x^2-y^2}$.
In Fig.~\ref{fig:crystal_interorbital}(b), we report the aligned energy diagrams for the 5-band model of the $1T$ and $1H$ phases, as well as for the 1-band/4-band model for the $1H$ polymorph.
The $d'_{z^2}$ WF is much lower in energy in the 1-band model, with an energy gain of $2.15$~eV compared to the $t_{2g}$ states of the $1T$ phase. 
On the other hand, the $d'_{xy}$ and $d'_{x^2-y^2}$ in the 4-band model are much higher in energy,  $3.27$~eV above the $d'_{z^2}$ state.

From the discussion above, we conclude that the dominant effect for the calculated trend in Fig.~\ref{fig:relative} is the inter-site hybridization between Wannier functions with different orbital character.
While the ligand field gives a small contribution, estimated in the 5-band model, its role is mostly an indirect one, i.e. producing different low-energy triplets in the two phases. 
\begin{figure*}[t]
\centering
\includegraphics[width=17cm]{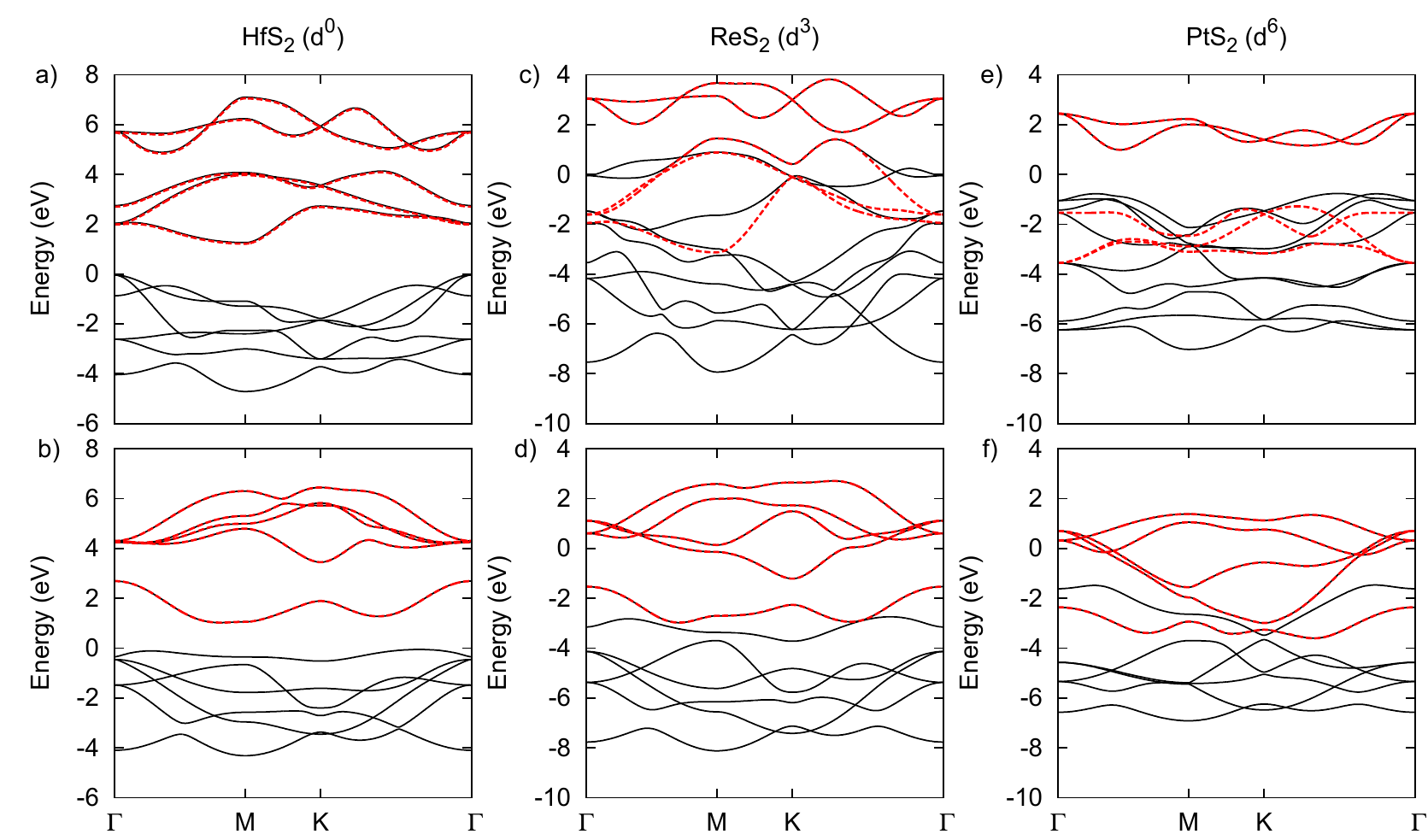}
\caption{\label{fig:evolution} Band structure calculated from first principles of the transition metal dichalcogenides (a)-(b) $1T$ and $1H$ HfS$_2$,(c)-(d) $1T$ and $1H$ ReS$_2$, and (e)-(f)  $1T$ and $1H$ PtS$_2$. The Fermi level is set to zero in all panels. The red dashed lines correspond to the band structure in the 5-band model.}
\end{figure*}

We would like to point out that the discussion here only applies to the relative stability of the ideal, undistorted $1T$ and $1H$ polymorphs.
Since TMDs are known to be subject to various lattice instabilities, we will briefly discuss further complications that can affect the energetics. 
Materials with $d^0$ occupations are either insulators or semimetals.
They are in general stable, with the notable exceptions of $1T$-TiSe$_2$, that undergoes an unusual insulator to insulator CDW transition \cite{sugawara2015unconventional}, and $1T$-TiTe$_2$, where a semimetal to semimetal CDW transition was recently observed in the limit of a monolayer but not in thicker samples \cite{chen2017emergence}.
This is associated with a small energy gain of a few meVs and does not affect significantly the overall relative stability.
In the $d^1$ family, the $1T$ and $1H$ (or $2H$) polymorphs are both observed experimentally and are subject to various forms of charge and spin instabilities \cite{manzeli_2d_2017, wilson1974charge, wilson1975charge, rossnagel2011origin, castro_neto_charge_2001, van2018competing, guller2016}. 
The corresponding energy gains are also of a few tens of meVs, but not necessarily negligible since the energy difference between the ideal $1T$ and $1H$ phases is very small (for TaS$_2$, the $1H$ phase is $62$~meV lower in energy).
It has been suggested that in specific cases these subtle effects might change the relative stability of the polymorphs \cite{isaacs2016, calandra2018}.
In that case, it is clear that a more careful treatment of electron correlations is needed to make a precise prediction.
While $1H$ TMDs with $d^2$ occupation are insulating and stable, the corresponding $1T$ materials are also found in nature in a distorted phase with a doubled unit cell \cite{chhowalla_chemistry_2013, whangbo1992analogies}.
The corresponding distorted $1T$ phase, dubbed $1T'$, has an energy much lower than the ideal $1T$ but is still unfavourable compared to the ideal $1H$ for all materials with the exception of WTe$_2$ \cite{santosh2015phase, duerloo2014structural}. 
In Fig.~\ref{fig:relative}, we see that for the $d^3$ case, the $1H$ phase is still slightly lower in energy for $4d$ and $5d$ disulfides. 
However, the corresponding materials, such as ReS$_2$, are most stable in a distorted $1T$ phase \cite{kertesz1984octahedral}, characterized by a $2 \times 2$ unit cell, and a large energy gain upon distortion \cite{tongay2014monolayer, choi2018origin}. 
The metallic $1H$ phase is not observed experimentally in any of the $n_d=3$ materials, and is predicted to be thermodynamically unstable \cite{tongay2014monolayer}.

We close this section by stressing that, although it can explain the trends of Fig.~\ref{fig:relative}, the ligand field/interorbital hybridization argument does not help understanding the higher stability of the $1T$ phase in the $d^0$ case.
It is expected that the electrostatic repulsion between the chalcogen atoms, that should be minimized in an octahedral coordination, plays an important role \cite{huisman1971trigonal}.
It is also possible that differences in energies of the $p$ bands favour the $1T$ polymorph. 
Estimating these effects is however outside of the scope of the present work.
\section{Trends accross the periodic table}\label{sec:trend}
\begin{figure*}
\centering
\includegraphics[width=17cm]{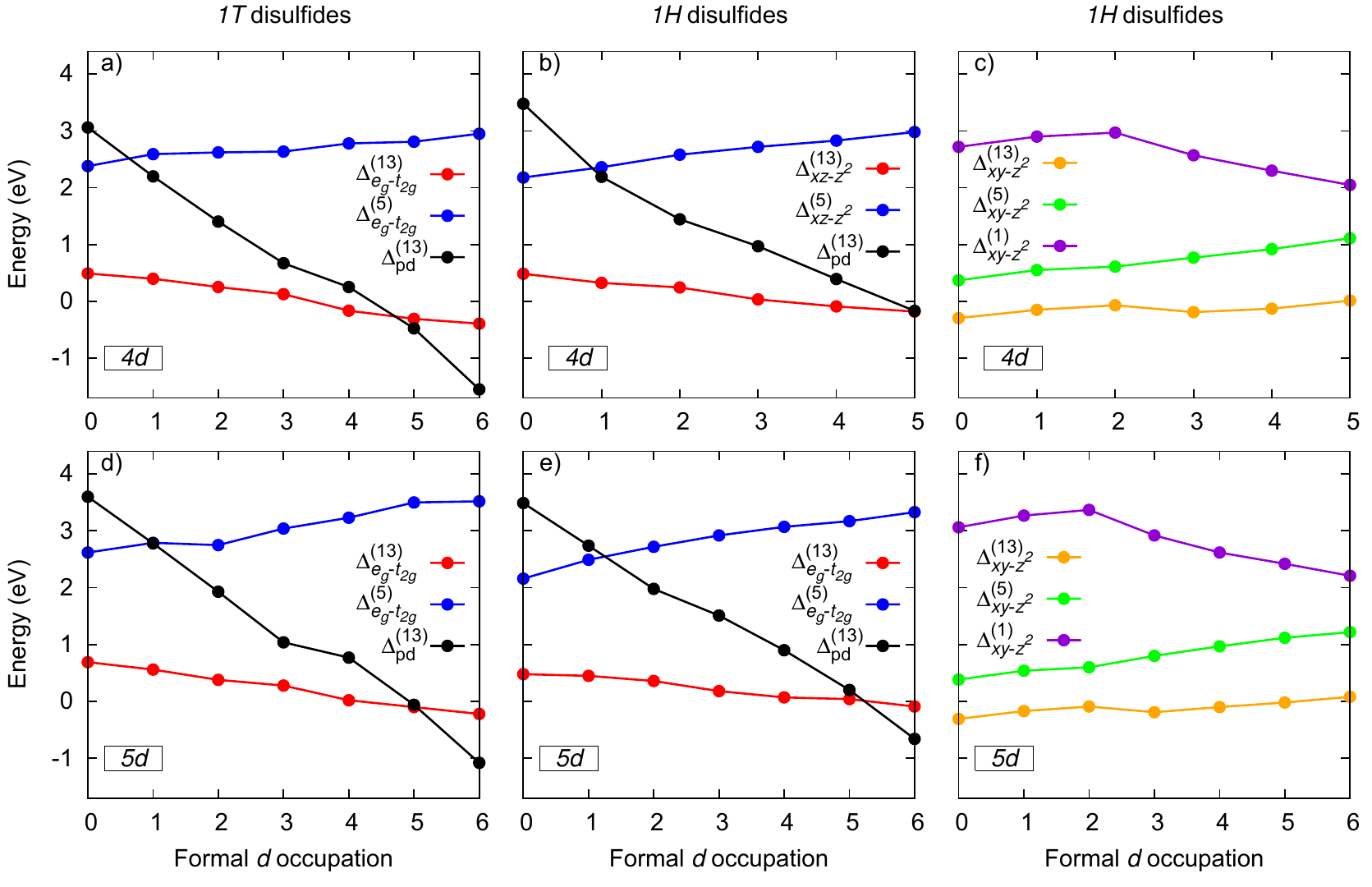}
\caption{\label{fig:trend_4d5d} Evolution of the calculated ligand field parameters as a function of the column, in the periodic table, of the transition metal for (a)-(c) $4d$ $1T$ and $1H$ disulfides, and (d)-(f) $5d$ $1T$ and $1H$ disulfides. }
\end{figure*}
So far, we have focused on the specific case of TaS$_{2}$.
In fact, as we shall discuss, because of the universality of the band structures of TMDs, the reasoning applies to the entire family of materials. 
As we shall see, the differences between materials are mostly quantitative, i.e. the calculated crystal field parameters vary smoothly across the periodic table and follow trends that can be understood with simple chemical intuition.
Reporting all the calculated parameters for all materials would not be particularly illuminating.
Therefore, we have chosen a few selected parameters and will discuss the evolution of those parameters in the following.

\begin{figure}
\centering
\includegraphics[width=8.5cm]{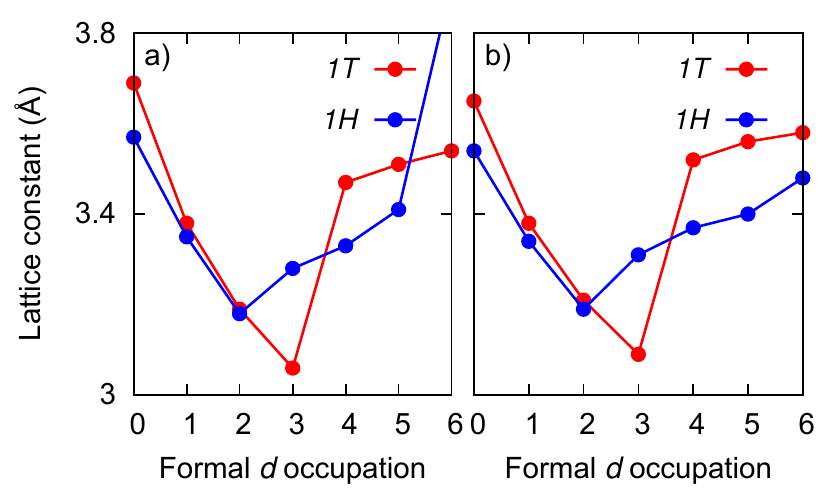}
\caption{\label{fig:alat} Evolution of the calculated lattice constant as a function of the column, in the periodic table, of the transition metal for (a) $4d$ $1T$ and $1H$ disulfides, and (b) $5d$ $1T$ and $1H$ disulfides.}
\end{figure}

We first consider the effect of changing the transition metal atom, while keeping the chalcogen to be sulfur.
In Fig.~\ref{fig:evolution}, we show the band structure of a few $5d$ transition metal disulfides in both polymorphs.
It is clear that they are qualitatively the same, the main difference being the position of the Fermi level. 
As the column of the transition metal in the periodic table increases, the energy separation between the $d$-like and $p$-like bands decreases. 
For $n\leq 3$, the two manifolds overlap, so that disentanglement is necessary to derive an effective 5-band $d$-only model. 
In Fig.~\ref{fig:evolution}, the disentangled $d$ bands are shown with red dashed lines. 
For the $1T$ case, the disentangled $d$ bands do not match the DFT bands, indicating hybridization with the $p$-like bands (e.g. avoided crossings). 
However, the resulting disentangled $t_{2g}$ bands look qualitatively similar in all cases.
The narrower bandwidth in the $d^6$ case (PtS$_2$) is not related to the entanglement with the $p$ bands, but to a larger lattice constant (see Fig.~\ref{fig:alat}).
For the $1H$ polymorph, it is the low-energy $d'_{z^2}$ band that overlaps in energy with the $p$ bands. 
In that case, the resulting disentangled bands match perfectly the DFT bands, meaning the absence of hybridization between these bands.
In Fig.~\ref{fig:evolution}, we have not plotted higher-energy bands. 
For $d^0$ disulfides, there is actually some overlap between the top of the $d$ bands and the bottom of the higher-energy bands, so that disentanglement is required to build the $d$, $pd$ and $spd$ models.

In Fig.~\ref{fig:trend_4d5d}(a) and (c), we plot the calculated $t_{2g}-e_g$ splitting in the 13-band and 5-band models, as well as the charge-transfer energy $\Delta_{pd}^{(13)}$ in the 13-band model, for MS$_2$, with M belonging to the $4d$ and $5d$ rows of the periodic table.
The band structures of the corresponding materials are shown in the Supplementary Information \cite{suppl}. 
The charge-transfer energy is defined as the average difference of on-site energies between the $d$ and $p$ Wannier functions, i.e. $\Delta_{pd}^{(13)} = \bar{\epsilon}_d^{(13)}-\bar{\epsilon}_p^{(13)}$.
We have taken the 13-band $spd$ model as representative of the bare electrostatic crystal field, so that the disentanglement of the high-energy bands is not necessary.

Figs.~\ref{fig:trend_4d5d} (a) and (c) show that, for both $4d$ and $5d$ disulfides, the crystal field in the 13-band model decreases as one moves to the right of the periodic table. 
This trend can be explained with chemical considerations, as the electronegativity increases as one moves to the left, favouring more ionicity and therefore a larger electrostatic contribution to the crystal field.
As one can observe in Fig.~\ref{fig:trend_4d5d}, the charge-transfer energy $\Delta_{pd}$ decreases as one goes to the right of the periodic table. 
Again, this trend can be understood from electronegativity considerations and is consistent with the maximal electrostatic contribution to the crystal field for $d^0$ TMDs.
For late-groups TMDs, we observe that the charge-transfer energy is negative. 
Consistently, the crystal field splitting in the $spd$ model is reversed for those materials.  
Note that for these late-groups TMDs, because of the small charge-transfer energy, hybridization is so large that the $d$ bands contain actually about $50$ percents of ligands' $p$ contribution.

While the electrostatic contribution to the crystal field is expected to decrease with decreasing charge-transfer energy, the opposite trend is anticipated for the ligand field since a small charge-transfer energy favours stronger hybridization.
The total ligand field, i.e. that calculated in the 5-band $d$-only model, is the sum of the bare electrostatic crystal field plus the contribution from hybridization with various ligand states.
Hence, the trend for $t_{2g}-e_g$ splitting in the 5-band model is controlled by the competition between opposite trends. 
For both the $4d$ and $5d$ cases, it appears that the hybridization trend dominates so that the total splitting increases for later-column materials. 

For the $1H$ polymorph, as shown in Fig.~\ref{fig:trend_4d5d}(b) and (e), the calculated trends are analogous to those in the $1T$ polymorph and the same logic applies. 
For both $4d$ and $5d$ disulfides, as one moves to the right in the periodic table, the $\Delta_{xz-z^2}$ splitting decreases in the $spd$ model and increases in the $d$ model. 
In the $4d$ case, we have not included $1H$-PdS$_2$ ($4d^6$) in the trend, because, as mentioned above, the electronic structure is qualitatively different.

In order to demonstrate the universality of the argument for the relative stability of the $1T$ and $1H$ phases, we study the $\Delta_{xy-z^2}$ splitting for the $4d$ and $5d$ disulfides.
In Fig.~\ref{fig:trend_4d5d}(c) and (f), we report the calculated splittings in the 13-band, 5-band and 1-band/4-band models. 
In the 13-band model, the splitting is consistently small and negative, except for $d^6$ TMDs because of the inverted charge-transfer energy.
In the 5-band model, the splitting increases linearly as $n$ increases due to the larger covalency.
On the other hand, in the 1-band model, the lowering of energy of the $d'_{z^2}$ WF is non-monotonous as a function of $n$.
It is interesting to notice that the maximum splitting  $\Delta_{xy-z^2}^{(1)}$ corresponds to the maximum filling of the low-energy state while keeping higher-energy states empty (i.e. $n=2$). 
This suggests that the lattice relaxes is such a way to maximize the $\Delta_{xy-z^2}^{(1)}$ splitting for the energy gain to be optimal.
Indeed, Fig.~\ref{fig:alat} shows that the calculated lattice constant for $1H$ disulfides follows the same trend.
The lattice constant is minimal at $n=2$ in order to increase intersite hoppings, while for the $1T$ case the minimum of the lattice constant is at $n=3$, corresponding to half-filled $t_{2g}$ bands. 
It is interesting to note that materials with the largest ligand field splittings $\Delta_{xy-z^2}^{(5)}$ do not exhibit the largest splittings $\Delta_{xy-z^2}^{(1)}$ when interorbital effects are included, confirming that the ligand field alone plays a minor role in stabilizing the $1H$ polymorph in $d^1$ and $d^2$ TMDs.

\begin{figure*}[t]
\centering
\includegraphics[width=17cm]{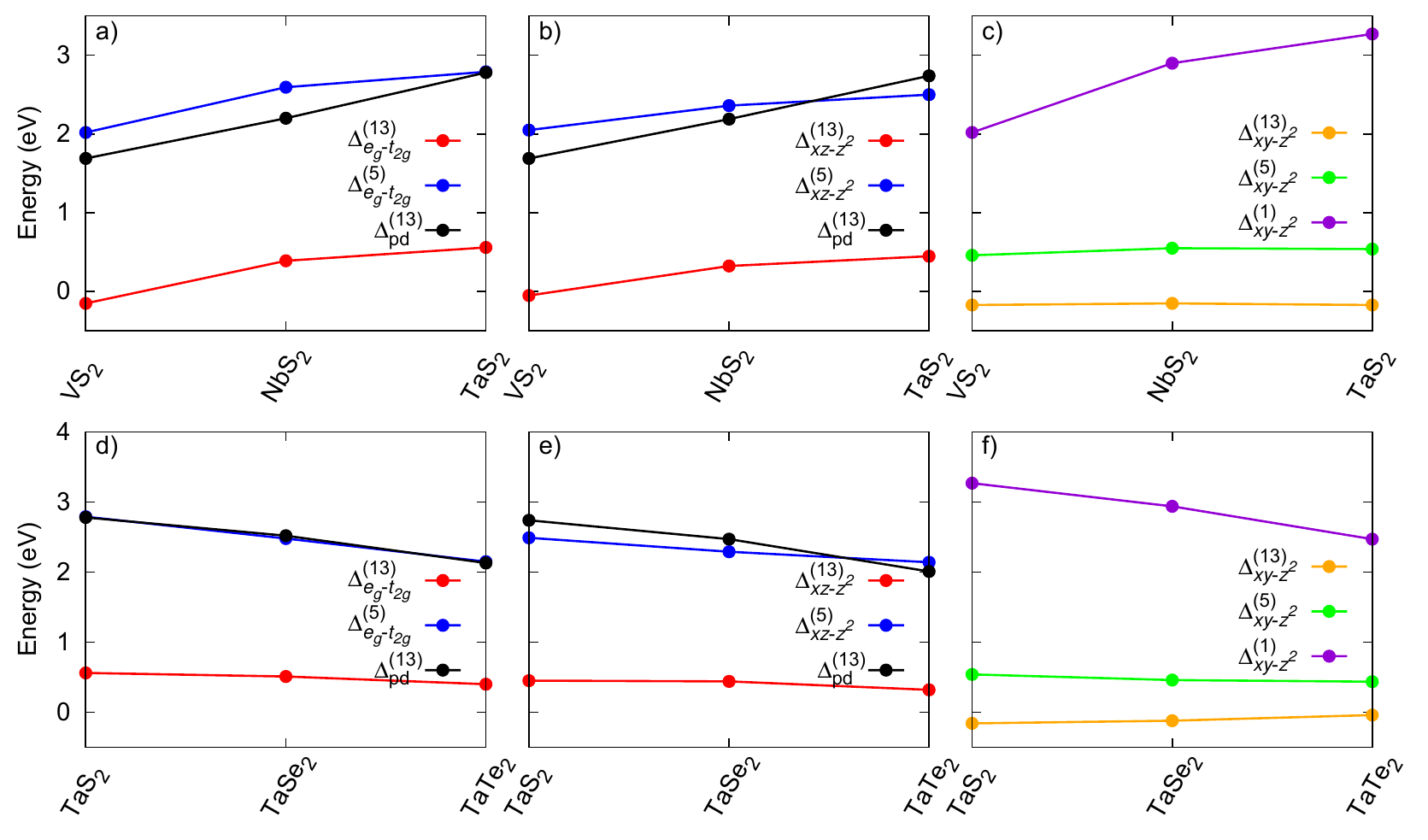}
\caption{\label{fig:trendtax2} (a-c) Evolution of the calculated ligand field parameters as a function of the row, in the periodic table, of the transition metal for $1T$ and $1H$ $d^1$ disulfides. (d-f) Evolution of the calculated ligand field parameters as a function of the chalcogen atom (X =S, Se, Te) for $1T$ and $1H$  TaX$_2$.}
\end{figure*}
In Fig.~\ref{fig:trendtax2}(a)-(c), we show the effect of changing the row of the transition metal atom in the periodic table, by considering $3d^1$ (VS$_2$), $4d^1$ (VS$_2$) and $5d^1$ (TaS$_2$) disulfides. 
Again, the calculated changes of the charge-transfer energy and the splittings in the $spd$ model follow trends that can be understood from electronegativity considerations.
The ligand field contribution to the splitting (i.e. difference of splittings between the $spd$ and $d$ models, not shown here) are almost constant, because the effect of a smaller charge-transfer energy in $3d^1$ materials is compensated by smaller hoppings, since the $3d$ electrons are more localized.
In Fig.~\ref{fig:trendtax2}~(c), we observe that the ligand field splitting $\Delta_{xz-z^2}^{(5)}$ is nearly constant for the three materials, but that the interorbital effects are larger for the $5d$ case.

In Fig.~\ref{fig:trendtax2}~(d)-(f), we summarize the effect of changing the chalcogen atom by considering TaS$_2$, TaSe$_2$, and TaTe$_2$. 
Again, the trend for the charge-transfer energy and crystal field splitting in the 13-band model follows what one can anticipate from simple chemistry considerations, as the electronegativity of the chalcogens decreases with increasing the row number in the periodic table, and is therefore the smallest for tellurium.
The somewhat smaller charge-transfer energy for TaTe$_2$ does not lead to increased ligand field splittings, because it is compensated by smaller hoppings due to a larger Ta-Te distance ($d=2.81$~\AA).
We note the trends for the energy splittings in the 5-band and 13-band models are similar, meaning that the trend for the overall splitting is controlled by the electrostatic effects.
\section{Relation to previous work}\label{sec:previous}
The question of the ligand field stabilization of the $1H$ (or $2H$ in the case of bulk materials) TMDs is an old one that goes back to the early days of research on layered dichalcogenides.
Therefore, before closing this paper, we wish to put our study in perspective with previous work.

The ligand field argument was put forward because of the discovery of stable $2H$ $d^1$ and $d^2$ TMDs, and still often appears in the recent literature.
Originally, there has been controversy regarding the alignement of the singlet state and low-energy $d_{xy}/d_{x^2-y^2}$ doublet \cite{goodenough1968band, huisman1971trigonal}.
Goodenough \cite{goodenough1968band} suggested a ligand field model with the $d_{z^2}$ singlet higher in energy than the $d_{xy}/d_{x^2-y^2}$ doublet.
In his model, the semiconducting character of $d^2$ TMDs such as MoS$_2$ is explained by the SOC-induced  splitting of the $d_{xy}/d_{x^2-y^2}$ doublet into $m_j=\pm 3/2$ and $m_j=\pm5/2$ singlets.
By considering both a simplified point-charge model and molecular-orbital calculations, Huisman \textit{et al.} \cite{huisman1971trigonal} suggested that, while electrostatic effects lead to a singlet higher in energy, hybridization with the ligands should reverse the ordering, in agreement with what we have found with our \textit{ab initio} Wannier-function approach.  
While Huisman \textit{et al.} estimated a ligand field stabilization for the trigonal prismatic coordination and suggested a simple picture for the electronic structure with a low-energy band derived from the $d_{z^2}$ state, Mattheiss \cite{mattheiss1973} showed how the actual band structure is more complex and stressed the role of intersite hopping effects in splitting the $d$ bands into a one-below-four pattern.
Mattheiss also estimated the ligand field splitting of the $d$ electrons, for MoS$_2$ and TaS$_2$, by fitting the $d$-like bands to a tight-binding model.
Surprisingly, the $\Delta_{xy-z^2}$ splitting of $0.04$~Ry ($\approx 0.544$~eV) for trigonal prismatic TaS$_2$ is in almost perfect agreement with our finding of $0.54$~eV.
Such agreement is likely accidental, as other features of the reported ligand field diagrams differ significantly from our results.
For instance, the $\Delta_{xz-z^2}$ splitting of $\sim 1.7$~eV is significantly smaller than what we have found ($2.5$~eV).
More importantly, the alignment between the $1T$ $t_{2g}$ states (which he discusses in terms of $d_{z^2}, d_{xy}$ and $d_{x^2-y^2}$ states, with $z$ oriented in the out-of-plane direction) and the $2H$ $d_{z^2}$ state is inverted compared to our results. 
Also, the reported splitting inside the $t_{2g}$ shell ($\sim 0.2$~eV) is significantly smaller than the value we obtained ($0.74$~eV) using the same coordinate system.

In this work, by systematically investigating the ligand field splittings across the family of materials, we have come to the conclusion that the ligand field does indeed have a stabilizing effect for $1H$ $d^1$ and $d^2$ TMDs, because the singlet $d_{z^2}$ state is lower in energy than the $1T$ $t_{2g}$ states for all materials considered. 
However, our quantitative calculations also show that this effect is fairly small (compared to the bandwidth or even compared to SOC-induced lifting of degeneracies), so that band structure effects are dominant and lead to a $d'_{z^2}$ Wannier function much lower in energy when interorbital hybridization is taken into account.

\section{Conclusion}\label{sec:conclusion}
In conclusion, using a modern Wannier-function-based methodology, we have revisited the problem of the relative stability of the $1T$ and $1H$ phases in TMDs by estimating crystal and ligand field parameters for a broad range of materials.
Our results show that the ligand field alone plays only a small if any role in determining the most stable phase, because the ligand field splitting of the low-energy triplet in the $1H$ phase is not large, and because the low-energy $d_{z^2}$ singlet state is found to be close in energy to the $t_{2g}$ triplet of the $1T$ phase.
This allowed us to conclude that intersite effects are dominant, so that the role of the ligand field is mostly an indirect one: giving rise to low-energy triplets with different orbital character in the two polymorphs.
We have also found that, because of the universality of the band structure, the variation of the calculated parameters vary smoothly across the family of materials and follow trends that can be understood using simple chemistry arguments.
Finally, our calculations show that the total ligand field splitting of the $d$-like states in TMDs arises from various contributions, i.e. from electrostatic repulsion effects and from the hybridization with the ligands' $s$, $p$ and $d$ states, that are all of a similar magnitude.
Therefore, simplified models, considering for instance only $pd$ bonding, should not be quantitatively correct.

A remaining question is that of the higher stability of the $1T$ phase for group $IV$ TMDs, that are characterized by empty $d$ bands. 
Quantifiying the effect of the repulsion between ions in the two coordinations would be an interesting next step in further elucidating the origin of the occurrence of different phases in this family of materials.
\section*{Acknowledgements}
We acknowledge funding by the European Commission under the Graphene Flagship (Grant agreement No.~696656). 
First-principles calculations were performed at the facilities of Scientific IT and Application Support Center of EPFL.
\bibliographystyle{apsrev4-1}
\bibliography{crystalfield}
\end{document}